\documentclass{aa}
\usepackage{graphicx}
\usepackage{txfonts}
\usepackage{hyperref}
\hypersetup{colorlinks = true, citecolor = {blue}, urlcolor= {blue}}

\renewenvironment{thebibliography}[1]{\begin{oldthebibliography}{#1}\setlength{\parskip}{0ex}\setlength{\itemsep}{0ex}}{\end{oldthebibliography}}

\usepackage{diagbox}
\usepackage{siunitx}
\newcommand{\orcid}[1]{\href{https://orcid.org/#1}{\protect\includegraphics[width=8pt]{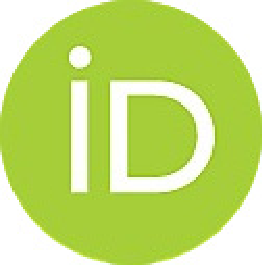}}}
\makeatletter
\renewcommand*\aa@pageof{, page \thepage{} of \pageref*{LastPage}}
\makeatother

\usepackage[dvipsnames]{xcolor}

\definecolor{lime}{rgb}{0.0, 0.0, 0.0}

\begin{document}
   
\title{Intermediate Mass Black Hole Binary Evolution in Nuclear Star Clusters: the effect of the stellar mass black hole population}

\author{Fazeel~Mahmood~Khan
\inst{1,2}\orcid{0000-0002-5707-4268}
\and
Peter~Berczik
\inst{3,4,5}\orcid{0000-0003-4176-152X}
\and
Margarita~Sobolenko
\inst{4,5}\orcid{0000-0003-0553-7301}
\and
Andreas~Just
\inst{6}\orcid{0000-0002-5144-9233}
\and
Rainer~Spurzem
\inst{6,7,8}\orcid{0000-0003-2264-7203}
\and
Kelly~Holley-Bockelmann
\inst{9,10}\orcid{0000-0003-2227-1322}
\and
Andrea V. Macci\`o
\inst{1,2}\orcid{0000-0002-8171-6507}
}

\institute{New York University Abu Dhabi, 
           PO Box 129188, Abu Dhabi, United Arab Emirates
           \and
           Center for Astrophysics and Space Science (CASS), New York University Abu Dhabi
           \and
           Nicolaus Copernicus Astronomical Centre Polish Academy of Sciences, ul. Bartycka 18, 00-716 Warsaw, Poland
           \and           
           Szechenyi Istvan University, Space Technology and Space Law Research Center, H-9026 Gyor, Egyetem ter 1. Hungary
           \and
           Main Astronomical Observatory, National Academy of Sciences of Ukraine, 27 Akademika Zabolotnoho St, 03143 Kyiv, Ukraine
           \and
           Astronomisches Rechen-Institut, Zentrum f\"ur Astronomie, University of Heidelberg, M\"onchhofstrasse 12-14, 69120 Heidelberg, Germany
           \and
           National Astronomical Observatories, Chinese Academy of Sciences, 20A Datun Rd., Chaoyang District, 100101, Beijing, China
           \and
           Kavli Institute for Astronomy and Astrophysics, Peking University, 5 Yi He Yuan Road, Haidian District, Beijing 100871, P.R. China
           \and
           Department of Physics and Astronomy, Vanderbilt University, Nashville, TN 37240, USA
           \and
           Department of Physics, Fisk University, Nashville, TN 37208, USA     }
   
\date{Received XXX / Accepted YYY}

\abstract
{}
{In this study, we investigate the dynamics of Intermediate-Mass Black Hole (IMBH) binaries within Nuclear Star Clusters (NSCs) that contain a population of stellar-mass black holes (BHs). We examine how these stellar and BH populations influence the dynamics of the IMBH binary and, in turn, how the evolving IMBH binary affects the surrounding stellar and BH populations. }
{We conduct high-resolution $N$-body simulations of NSCs constructed based on observational parameters from two local dwarf galaxies: NGC205 and NGC404. For the first time, we achieve a star particle mass resolution of $1\rm\;M_{\odot}$  and a BH mass resolution of $10\rm\;M_{\odot}$. This level of resolution is crucial for accurately modeling the collisional dynamics of these dense systems. } 
{Including stellar-mass BHs within the stellar population significantly influences the IMBH binary dynamics, nearly doubling the sinking rate and halving the merger time. During the initial phase of the inspiral, the IMBH binary disrupts both the stellar and BH cusps. However, the BH cusp quickly regains its steep slope due to its shorter relaxation time and continues to dominate the evolution of the IMBH binary, despite being much less massive compared to the stellar component. 

We uncover an interesting mechanism in which BHs first efficiently extract energy from the IMBH binary and then transfer this energy to the surrounding stars, allowing the BHs to spiral back toward the center of the system and restart the process.

Our results imply that, although stellar mass BHs are a minor component of a stellar population, they can significantly facilitate IMBH growth within NSCs via mergers. We also notice that these dense systems can potentially boost Intermediate Mass Ratio Inspirals (IMRIs) predominantly on radial orbits. }
{}


\titlerunning{IMBH binaries in NSCs}
\authorrunning{Khan~et.~al}
\maketitle

	

\section{Introduction}\label{sec-intro}

Intermediate-Mass Black Holes (IMBHs), with masses between $10^2$ and $10^6$ times that of the Sun, represent an elusive segment of the astrophysical black hole (BH) mass spectrum, which spans a vast range from approximately 10 solar masses to several billion solar masses \citep{Kormendy+13,graham+15}.

Local scaling relations between galaxies and their central BHs \citep{sturm24} suggest that IMBHs should exist at the centers of dwarf galaxies. These BHs can be identified either by the detection of gas accretion signatures or by dynamical studies that resolve the immediate vicinity of the BH, often referred to as its sphere of influence \citep{Reines2022,ask23}. Over the past decade, several IMBH candidates have been identified in dwarf galaxies through spectroscopic evidence of active galactic nuclei \citep[AGN;][]{Molina21}. Additionally, X-ray emission observed with the \textit{Chandra} X-ray Observatory have provided further support for the presence of active BHs in these small galaxies \citep{bald17}. Other indirect evidence includes optical variability in AGNs \citep{bald20} and tidal disruption events \citep{fre20,Yao2025}, both of which have been linked to the existence of IMBHs in dwarf galaxies. IMBHs with low accretion rates or those that are inactive, along with BHs in dwarf galaxies that are experiencing significant star formation, are challenging to detect, leaving the true occupation fraction of IMBHs in dwarf galaxies uncertain.

Many IMBH hosts also contain a dense central concentration of stars known as Nuclear Star Clusters (NSCs), which are among the densest stellar environments in the universe \citep{neu20}. NSCs typically have masses between $10^5$ to $10^7$ solar masses, with sizes ranging from a few to about 10~pc \citep{bok04,denbrok+2014,geo14,spe17}. NSCs with the highest occupation fractions ($\lesssim $ 90\%) are observed in galaxies with stellar masses around $10^9$ solar masses \citep{san19b}. These NSCs are believed to be assembled through the infall of globular clusters \citep{tre75,cap93,Antonini+12} as evidenced by their old stellar populations, although the existence of a young stellar population in some NSCs suggests that \textit{in situ} star formation \citep{milo04,ant15} may also play a role. The efficiency of either assembly mechanism depends on the mass of the host galaxy. In more massive galaxies, \textit{in situ} star formation is considered the dominant process, while in lower-mass galaxies, globular cluster inspiral appears to have a more critical role \citep{fah22}. 


Due to their extremely high central densities, NSCs have relatively short relaxation times, on the order of a few billion years \citep{mer09}. NSCs are also expected to harbor a population of compact objects, such as neutron stars (NSs) and BHs. Because these objects are more massive than the majority of the stellar population, they undergo mass segregation and sink toward the center of the cluster. Additionally, since the relaxation time is inversely proportional to the mass of the species, these compact objects are expected to relax much earlier than their stellar counterparts and form its own dynamical substructure. Within the influence radius of an IMBH, for example, the relaxed NS and BH population is likely to form a \citet[][BW]{Bahcall1976} cusp, with a characteristic $r^{-7/4}$ density profile.

Multiple IMBHs can accumulate in a dwarf galaxy nucleus through the dynamical friction-driven inspiral of massive star clusters containing IMBHs, or through mergers between dwarf galaxies. Once in the galactic center, these IMBHs can form a bound binary system, which may further evolve through stellar dynamical processes \citep{begelman+80} and eventually coalesce via gravitational wave (GW) emission. The emission of GWs from IMBH binary sources will be detectable up to high redshift by Einstein Telescope (ET) and Cosmic Explorer (CE) \citep{DiGiovanni2025}, their access to
low frequencies will make it possible to observe population of intermediate-mass black holes up to about $10^4$M$_{\odot}$ \citep{Sathyaprakashetal2012,Abacetal2025}\footnote{
\href{https://www.et-gw.eu/}{https://www.et-gw.eu/ , Einstein Telescope (ET)}}.
In the near future, the ESA/NASA space-borne GW observatory, Laser Interferometer Space Antenna (LISA) and the Chinese projects TianQin and Taiji\citep{Gongetal2021} will be able to detect IMBH mergers as well
 \citep{ama23,colpi24,McCaffrey2025} \footnote{\href{https://lisa.nasa.gov}{https://lisa.nasa.gov , LISA}}.

Recently, several studies have explored the binary IMBH evolution in merging NSCs \citep{ogi19,mukh23,mukherjee24} or in an isolated galaxy environment \citep{Khan_Holley-Bockelmann2021}. These studies conclude that IMBHs pair, form a binary system, and merge very efficiently in the dense NSC environment on a timescale ranging from about 100~Myr to approximately a Gyr. There can be a 10-fold difference in the merger timescales due to various structural and kinematic properties such as density and rotation of the host with respect to the initial orbit of inspiraling IMBH.   In the absence of a NSC in dwarf galaxies, IMBH binaries are shown to stall for longer than a Hubble time around a parsec separation \citep{bia19,khan+24,Partmann2024,zhou25}. 

\citet{mukh23} studied IMBH binary evolution in mergers between mass-segregated NSCs, and found a strong dependence on the IMBH binary mass ratio; orbital evolution is accelerated for high mass ratio IMBH binaries, while it is prolonged when the IMBHs have comparable masses. However, we note that with stellar particle masses of 14.5~M$_{\odot}$ and BH particle masses of 145~M$_{\odot}$, relaxation may be an issue. Another caveat for this study is that only the NSCs were modeled. \citet{Khan_Holley-Bockelmann2021} have shown that then the NSC-to-IMBH mass ratio $\leq 10$, stellar encounters originating from the bulge significantly contribute to shrinking the binary orbit. For mass ratios greater than 10, IMBH binary evolution is governed predominantly by the stellar encounters originating from the NSC, and contributions from the surrounding bulge can be ignored.  

We study IMBH binary evolution in observationally-motivated models of nearby NSC-embedded dwarf galaxies in which the NSC-to-IMBH mass ratio is greater than 10. For the first time in such studies, we achieve 1~M$_{\odot}$ resolution for the stellar component, which is crucial to capture the accurate dynamics of collisional systems such as an NSC. We also include a stellar mass BHs population (10~M$_{\odot}$ each) within the influence radius of the IMBH that constitutes 1\% of the total NSC mass \citep{pan19}. Our models are based on NGC205 and NGC404, because their structural, kinematic, and dynamical mass measurements are well-resolved within the radius of influence of the IMBH~\citep{ngu17,ngu18}. NGC205 and NGC404 are also ideal because the NSC-to-IMBH mass ratio is greater than 10, which allows us to focus only on the central NSC in unprecedented mass resolution.

The manuscript is organized as follows: Section~\ref{sec:nsc-models} contains a description of our models and numerical techniques, Section~\ref{sec:results} shows the results of IMBH binary evolution in NGC205 and NGC404 NSC models, and finally Section~\ref{sec:summary} presents a summary and concludes the findings of our study.

\section{NSCs Models and Their Stability} \label{sec:nsc-models} 

We build NSC models with parameters chosen for those observed in NGC205 \citep{ngu18} and NGC404 \citep{seth10,ngu17}. NGC404 has two NSCs in addition to a surrounding bulge. However, here we build a composite model of the two NSCs such that its profile matches that of the combined profile of the two NSCs up to 5~pc.  Our models also include a dark component in the form of stellar mass BHs distributed in a BW cusp \citep{Bahcall1976} within the sphere of influence of the central IMBH; the BH are initialized in a Dehnen profile \citep{deh93}  with $\gamma = 1.75$. Table~\ref{tab:paramnsc} shows the parameters of our initial NSC models. For the NGC205 NSC, the stellar component is constructed with 1.84~million particles, giving us a mass resolution of 1~M$_{\odot}$, whereas the stellar mass BH component has 1800~particles, achieving the desired mass resolution of 10~M$_{\odot}$. The NSC in NGC404 is more than twice as massive as that of NGC205, requiring 4.25~million particles for the stellar component and 4250~particles for the BH component to achieve the same mass resolution. Our models are mildly triaxial, which is a key to ensuring a sustained supply of stars on centrophilic orbits in a hard IMBH binary regime \citep{Baile15}.

To distinguish between the impact caused by the presence of the stellar mass BH component and the stellar distribution, we compare to models with the exact same initial stellar profile, but no stellar mass BHs. In the stellar-only models, we add the mass of the BH component to that of the stellar mass.


\begin{table*}
\begin{center}
\vspace{-0.5pt}
\caption{Initial NSCs parameters.} 
\begin{tabular}{ c c c c c c}
\hline
\hline
NSC Component & $N$ & Mass ($10^6$ M$_{\odot}$) & $R_{\rm eff}, R_{\rm infl} $ (pc) & $n$ or $\gamma$  & $b/a,c/a$ \\
\hline
\multicolumn{6}{c}{\textbf{NGC205}}\\
\hline
Stars & $1.84 \times 10^6$ & $1.84$ & $1.3$ & $1.6$ & $0.95,0.85$\\
BW Cusp & $1.8 \times 10^3$ & $0.018$ & $0.14$ & $1.75$ & $0.95,0.85$\\
IMBH & $1$ & $0.022$ & $0.14$ & --- & --- \\
\hline
\multicolumn{6}{c}{\textbf{NGC404}}\\
\hline
Stars & $4.25 \times 10^6$ & $4.25$ & $1.85$ & $0.65$ & $0.95,0.85$\\
BW Cusp & $4.25 \times 10^3$ & $0.0425$ & $0.35$ & $1.75$ & $0.95,0.85$\\
IMBH & $1$ & $0.027$ & $0.35$ & --- & --- \\

\hline
\end{tabular}
\begin{minipage}{\linewidth}
\smallskip
Column~1: NSC model and its components. Column~2: Number of particles. Column~3: Component mass in million solar masses. Column~4: Effective radius of NSC component or influence radius of IMBH in pc. Column~5: S\'ersic index for stellar component or Dehnen profile power law index for inner cusp. Column~6: Intermediate to major and minor to major axes ratios.
\end{minipage}
\label{tab:paramnsc}
\end{center}
\end{table*}

We utilize publicly accessible AGAMA software \citep{vasi19} to create our NSC components, adopting the Schwarzschild method to generate triaxial models. Our initial models are in dynamical equilibrium with the central IMBH. Figures~\ref{fig:stab205} and \ref{fig:stab404} show density (\textit{top panel}) and mass profiles (\textit{bottom panel}) of stellar and BH components for NGC205 and NGC404 NSC models, respectively, at the initial time. For the NGC205 model, the BH component in the form of a BW cusp dominates both the density and cumulative mass in central region but as we approach the influence radius of the IMBH (marked by the vertical brown line), the stellar component clearly starts to take over and as we approach the effective radius of the NSC (marked by the grey vertical line), the stellar component dominates in mass by two orders of magnitude. The mass enclosed inside the sphere of influence of the IMBH for the NGC404 model is similar for both the stellar and BH components. Again, as $r$ approaches the effective radius of the NSC, the stellar mass dominates by almost two orders of magnitude.

\subsection{Simulation codes}\label{codes}

We use two versions of our $N$-body code; $\phi-$GPU \citep{berczik+11,Berczik2013} and $\phi-$GPU hybrid, both employ a 4th order Hermite integration scheme to advance particles forward in time. Both codes also utilize individual block time-steps. $\phi-$GPU is a pure $N$-body code, calculating pairwise forces between particles resulting in O($N^2$) scaling with particle number. It also employs individual softening for each particle. We have used $\phi-$GPU extensively for large-scale $N$-body simulations of supermassive BHs in galaxy mergers \citep{Khan+15,kha16,sobo22,Khan2025LRD} as well as for IMBH binary evolution in nucleated dwarf galaxies \citep{Khan_Holley-Bockelmann2021}. 

The $\phi-$GPU-hybrid, as name suggests, employs a hybrid scheme to calculate the forces of each particle. It divides the particles into three categories: BHs, core, and halo particles. For both BH and core particles, forces are calculated in a pairwise scheme as in the case of $\phi-$GPU; however, for halo particles, it employs the Self-Consistent Field (SCF) method for force calculations \citep{hern1992}. We put the inner 10\% of particles as the core particles. This covers roughly all BH particles and stellar particles upto three influence radii of the IMBH. This is the region with the highest density, where the relaxation time is the shortest, which means that the two-body interactions should be resolved effectively. Well beyond the influence radius, the stellar and BH densities fall considerably, causing the relaxation time to become larger, and the particles are expected to effectively follow the global potential created by the mass distribution. We employ the $\phi$-GPU-hybrid code to evolve the IMBH binaries in the later phase of simulations that include both stellar and stellar-mass black hole components, once the binary reaches a `hard binary' separation, commonly defined as \citep{Merritt_2013}:

\begin{equation}
    a_{\mathrm{h}} \equiv \frac{G \mu}{4\sigma^{2}},
\end{equation}

where $\mu$ is the reduced mass of the binary and $\sigma$ is the velocity dispersion of the background distribution.

For the pairwise force calculation, we adopt an individual softening length of $10^{-4} ~\mathrm{pc}$ for the stellar and stellar-mass black hole particles. The forces between the IMBH pair are always computed with zero softening.
Although we employ particle-specific softening to suppress numerical divergences, its value is very small ($\sim 10^{-4}$ pc) retaining accurate close-encounter dynamics.
\subsection{Models Stability}\label{stability}
\begin{figure}
  \centering
  \includegraphics[width=0.98\hsize]{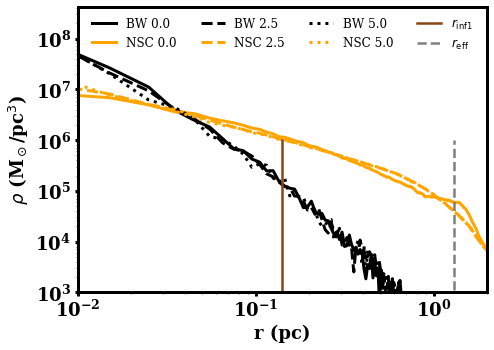}\\[1ex]
  \includegraphics[width=0.98\hsize]{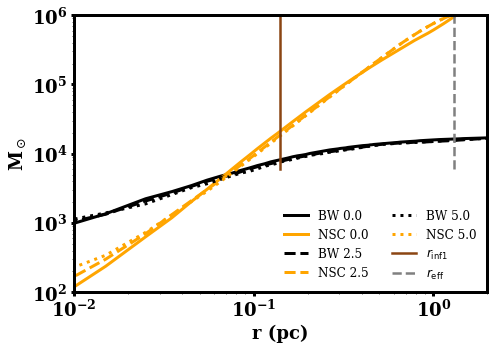}
  
  \caption{
  Volume density profiles of individual stellar (NSC) and BH (BW) components (\textit{top panel}) and cumulative mass profiles of the same (\textit{bottom panel}) for NGC205 NSC+BW model. The full lines represent initial profiles for the BH and stellar components, and dashed lines represent the same (in corresponding colors) at designated times in Myr. The brown vertical line represents the sphere of influence ($r_{infl1}$) of the IMBH for initial distribution, and the gray vertical line showcases the effective radius of the NSC.
  }
  \label{fig:stab205}
\end{figure}


\begin{figure}
  \centering
  \includegraphics[width=0.98\hsize]{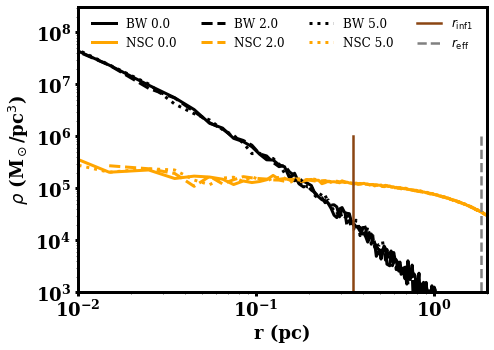}\\[1ex]
  \includegraphics[width=0.98\hsize]{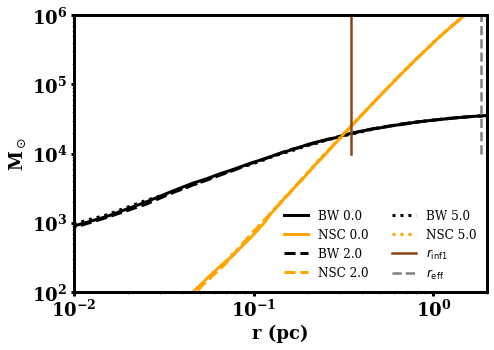}
\caption{
The same as in Fig.~\ref{fig:stab205}, but this time for the NGC404 NSC+BW model.
} \label{fig:stab404}
\end{figure}

Before invoking a secondary IMBH in our NSC models, we run them for stability. We used $\phi-$GPU to test the stability of our models. We evolve our models for upto 5~Myr. In our runs, an IMBH binary forms before 5~Myr, and the subsequent evolution of the system is greatly impacted by the IMBH binary presence. 

Figures~\ref{fig:stab205} and \ref{fig:stab404} show the initial and time-evolved density and cumulative mass profiles of the individual components of the NSCs for NGC205 and NGC404, respectively. Both density and cumulative mass profiles of individual components (stellar and BHs) appear to be reasonably stable throughout our stability analysis run for the two NSCs. 

\subsection{Secondary IMBH}\label{orbits}

We introduce secondary IMBHs with masses that are four times smaller than the primary IMBH in both models, this is $5.5\times10^{3}\rm\;M_{\odot}$ for NGC205 and $6.75\times10^{3}\rm\;M_{\odot}$ for NGC404. Only in the case of NGC205, we also introduce a BH in 40~times smaller with mass $5.5\times10^{2}\rm\;M_{\odot}$. NGC404 NSC has a high number of particles ($N= 4.25 \times 10^6$), and it becomes computationally very expensive to model the 1:40 mass ratio IMBH binary case due to longer dynamical friction times. These secondary IMBHs are initially placed at a separation of 1~pc from the primary IMBH, with velocities equal to half the local circular velocity. The initial separations are well outside the influence radii of the IMBHs and are close to the effective radius of NSCs. We assume that secondary IMBHs to such a separation can be delivered by either a merger with a smaller dwarf galaxy or by an inspiralling star cluster having a central IMBH.  

Direct $N$-body simulations have demonstrated the formation of IMBH in dense star clusters by the following mechanisms: several generations of mergers of stellar mass black hole binaries through gravitational wave emission \citep{Rizzutoetal2022,Rizzutoetal2023,ArcaSeddaetal2024}, or by stellar collisions building a growing very massive stars, which then collapses to an IMBH \citep{Vergaraetal2024,Vergaraetal2025}. In \citet{RantalaNaab2025} a combination of both processes leads even to the formation and merger of binaries of IMBH. All processes in the cited papers happen in a timescale of the order of 5 Myr.

\section{Results}\label{sec:results}
Here, we discuss the results of our $N$-body simulation suite for both NGC205 and NGC404 NSC models.

\begin{figure*}
     \centering
     \includegraphics[width=0.97\hsize]{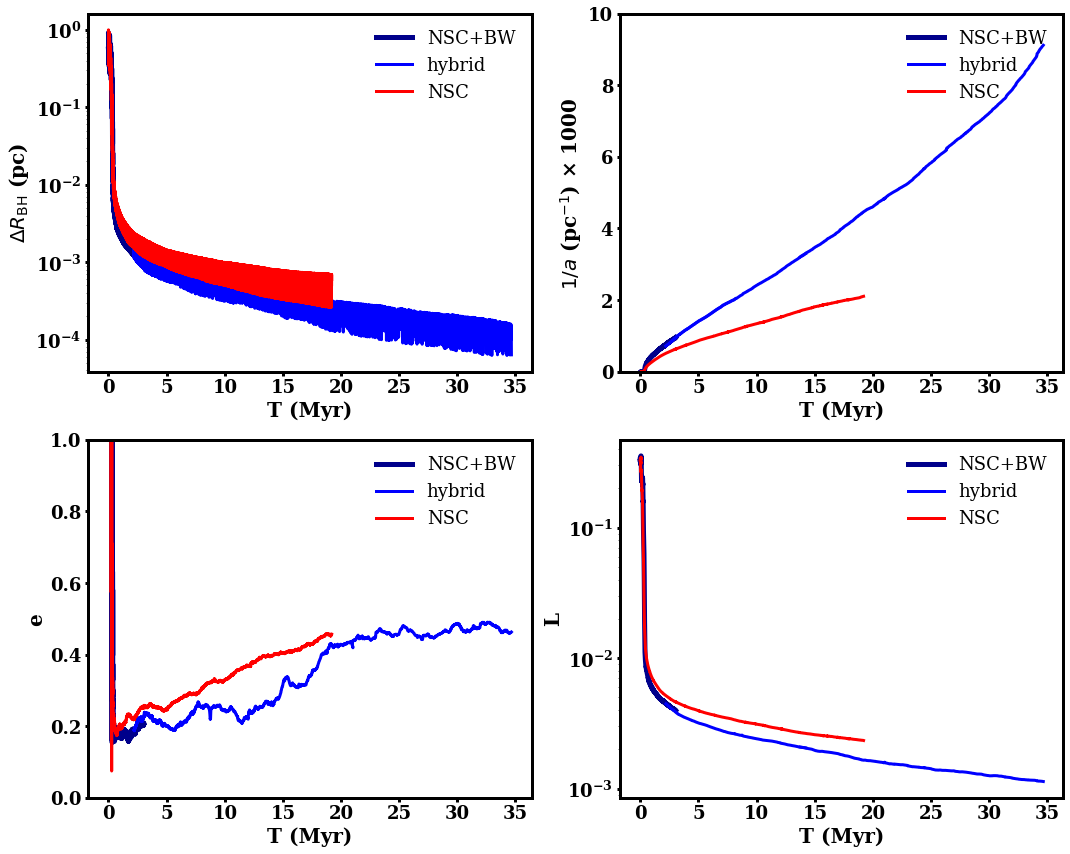}
    \caption{IMBH pair and binary evolution in NGC205 models with (blue colors) and without (red color) a central BH cusp for a 1:4 IMBH mass ratio. The IMBH binary evolution in simulation performed with $\phi-$GPU hybrid code is labeled as 'hybrid'.  \textit{Top left}: IMBHs orbit relative separation ($\Delta R_{\rm bh}$). \textit{Top right}: IMBHs orbit inverse semimajor axis ($1/a$). \textit{Bottom left}: IMBH binaries eccentricity ($e$). \textit{Bottom right}: Total angular momentum ($L$) of IMBH binaries in units of $1.64 \times 10^8$ $M{\odot}$~pc~km/s.}
    \label{fig:img1}
\end{figure*}

\begin{figure}
    \centering
  \includegraphics[width=0.98\hsize]{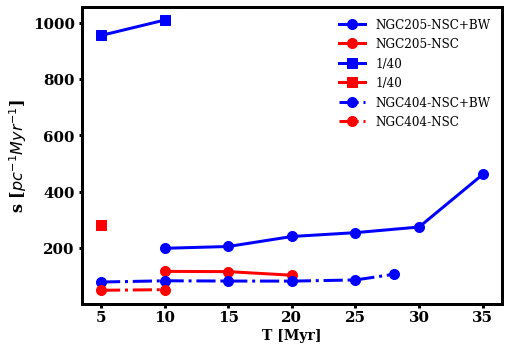}\\[1ex]
    \caption{IMBH binary hardening rates $s$ at various times for all of our runs.}
    \label{fig:s0p25}
\end{figure}


\subsection{NGC205: IMBH binary evolution} \label{sub:ngc205}

We studied IMBH binary dynamics in NGC205 NSCs (with and without the BH component) with secondary IMBH having masses that are 4 and 40~times smaller than that of the primary IMBH.

\subsubsection{NGC205: IMBH binary (1:4) evolution} \label{subsec:simulation1to4}

We start with the results of IMBH pair evolution in NGC205 with a 1:4 mass ratio run -- where the secondary is 4~times less massive than the primary. The evolution of the IMBH pair for this model, both with and without a BH component, is shown in Fig.~\ref{fig:img1}. IMBHs' separation (\textit{top left panel}) initially shrinks rapidly due to dynamical friction as the IMBHs become bound ($T_{\rm b} \approx 0.4$~Myr) and subsequently lose orbital energy via gravitational slingshots, shrinking their separation well below 0.01~pc in the first Myr of our simulation for both cases. Understandably, the separation shrinks quickly for a model with a BH component, as it provides additional background density for the dynamical friction phase and also for the three-body scattering phase when the IMBH binary separation shrinks below 0.1 pc. The shrinking of the binary, from its initial formation to the stage of hard binary formation, occurs over roughly a dynamical timescale -- the timescale during which particles interact with the binary. For the NGC205 models, this timescale is approximately 1~Myr at a radius of 0.1~pc, both for models with and without BHs, since around $r = 0.1$~pc, the stellar and BH densities happen to become comparable.

Later, a more steady phase of evolution governed by three-body scattering of stellar and BH particles is witnessed after a hard binary \citep{Merritt_2013} forms around an inverse semimajor axis $1/a$ value of 500~pc$^{-1}$ (\textit{top right panel}). IMBH binaries form with a smaller value of eccentricity $e \approx 0.2$ for both cases, which is growing to a value of $e \approx 0.4$ during our run (\textit{bottom left panel} of Fig.~\ref{fig:img1} ). The \textit{bottom right panel} shows the time evolution of the IMBH binary angular momentum. We estimate the IMBH binary hardening rate $s= d/dt(1/a)$ at various times by fitting a straight line to the evolution of the inverse semimajor axis in the hard binary regime with a 5~Myr interval. Figure~\ref{fig:s0p25} shows the time evolution of the hardening rate for both cases: red filled circles with the solid line for the NSC+BW model and dashed line for the NSC model alone. The IMBH binary hardening rate is considerably higher (approximately 2-3 times) in the model with the BW cusp of stellar BHs when compared to the NSC model without a BH component. Also, both hardening rates show different trends. The $s$ increases with time in the model with the BH component, whereas for the stellar component only model, it remains more or less constant with a slight decrease during the last interval witnessed in our simulation. 


\begin{figure}
\centering
  \includegraphics[width=0.98\hsize]{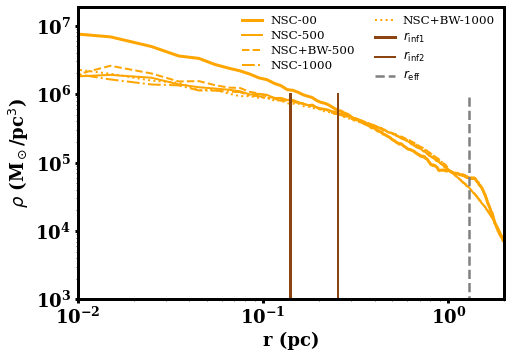}\\[1ex]
  \includegraphics[width=0.98\hsize]{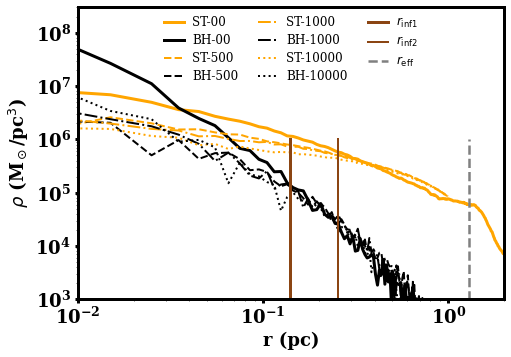}
\caption{ Density profiles for NGC205 models with 1:4 mass ratio. 
\textit{Top panel}: Density profile of stellar components in NSC+BW and NSC models. The 00 line is the initial stellar profile, which is the same for both models. Thick and thin brown vertical lines are influence radii at the start and end of our NSC+BW model run, respectively. The different lines are density profiles of the designated models at the labeled $1/a$ value of the IMBH binary. 
\textit{Bottom panel}: Density profile of stellar (ST) and black hole (BH) components in NSC+BW model. The full lines are initial profiles and dashed/dotted lines are profiles at designated values of $1/a$ for stellar (orange color) and black hole component (black color).} 
\label{fig:205dens-stars+bh}
\end{figure}

\begin{figure}
\centering
  \includegraphics[width=0.98\hsize]{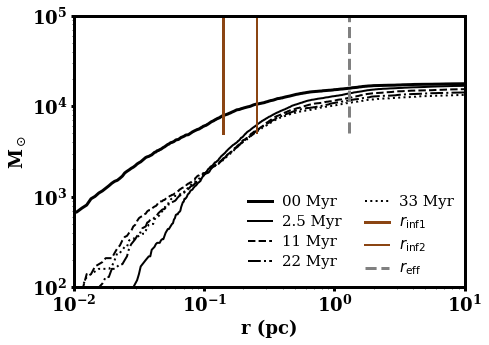}\\[1ex]
  \includegraphics[width=0.98\hsize]{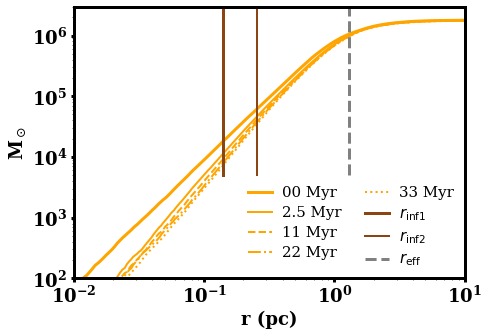}
\caption{
Cumulative mass profiles (in solar masses) of the BHs (\textit{top panel}) and stellar distribution (\textit{bottom panel}) for the NGC205 model with BH component for 1:4 mass ratio at various times during IMBH binary evolution. Thick and thin brown vertical lines are the initial and final values of the influence radius, respectively, and the gray vertical line is the initial effective radius of NSC. 
} \label{fig:bbh205}
\end{figure}

In order to understand the reason behind the increased hardening rate for the BH component model, we analyze various properties of our models during the late stages of IMBH binary evolution. We first look up at the mass volume density profiles of the stellar component (only) in both models, with and without the BH population (see \textit{top panel} of Fig.~\ref{fig:205dens-stars+bh}) at various values (labeled in figure) of $1/a$. The density profiles for a particular $1/a$ value are very similar for the cases at influence radii (initial and final values). We notice a slightly higher density value for the NSC+BW model as we move closer to the center ($r \approx 0.01$~pc). This slight increase is not sufficient to explain a factor $2-3$~ of the hardening rate difference in the two models. Next, we plot the mass volume density of both the stellar and the BH components for the NSC+BW model (see \textit{bottom panel} of Fig.~\ref{fig:205dens-stars+bh}). Here, we notice that the density of the stellar component keeps on decreasing in the central part, as is expected from the core scouring by IMBH binary via three-body scattering \citep{mer06,berczik+06,khan+11,khan+12a,Gualandris+12,kha16,rantala+18} resulting in partial disruption of NSC as is witnessed in \citet{Khan_Holley-Bockelmann2021}. In contrast, the density profile of the BH component tends to become steeper after initial core scouring. The same trend is translated to cumulative mass profiles evolution (Fig.~\ref{fig:bbh205}) for BH and stellar components, where we see that the BH component steepens after initial core scouring by the IMBH binary, and stellar mass continues to decline inside the influence radius.

\begin{figure}
\centering
     \includegraphics[width=0.98\hsize]{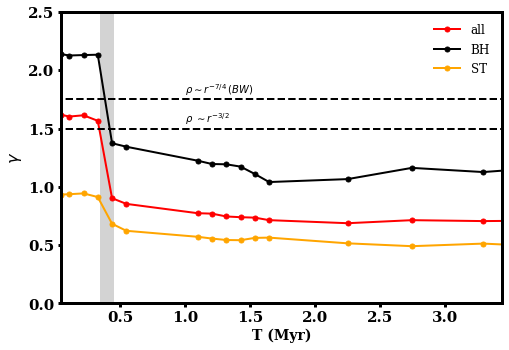}
\centering
     \includegraphics[width=0.98\hsize]{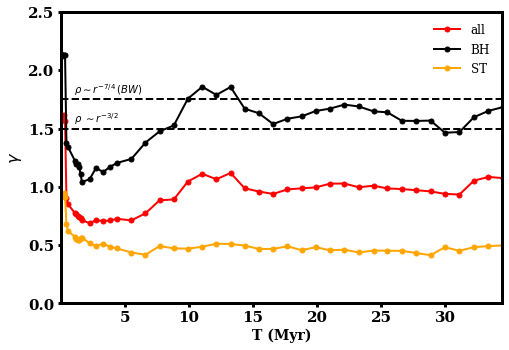}
\caption{
{ Evolution of the logarithmic density slope parameter $\gamma$ for NGC205 with 1:4 mass ratio. $\gamma$ was calculated by fitting and averaging over the nine snapshots in a timespan of $\approx 0.1$~Myr. The errors of the averaging are less than one per cent. The colour coding represents three different components: BHs (black), stars (orange), and all particles (red). \textit{Top panel} shows the $\gamma$ for the first $3.5$~ Myr of the evolution, whereas \textit{bottom panel} shows the same for full-time evolution. Grey zone in the \textit{top panel} indicates the IMBH binary formation to a hard binary transition time. Dashed horizontal lines show $\rho(r)$ relations for various power indices, including the BW cusp.}
}
\label{fig:power-index-205}
\end{figure}

We plot the slope of the inner cusp during IMBH binary evolution (see Fig.~\ref{fig:power-index-205}) for the density profile as a function of time for individual species and also for the combined mass of stars and BHs. 
We can see from the \textit{top panel} of the figure how the power index ($\gamma$) drops as the binary transitions from its formation to a hard binary regime, indicating flattening of density slopes in the center for both the stellar and BH distributions (as is evident from Fig.~\ref{fig:205dens-stars+bh}). As the IMBH binary evolves further in time (\textit{bottom panel}) in the hard binary regime, we notice that the profile of stellar component maintains roughly the same slope (shallow density profile), however, the density profile of the BH component evolves and approaches $r^{-7/4}$ and remains close to this as it is a steady state solution for the dominant mass component around a massive BH. 

\begin{figure}
\centering
     \includegraphics[width=0.98\hsize]{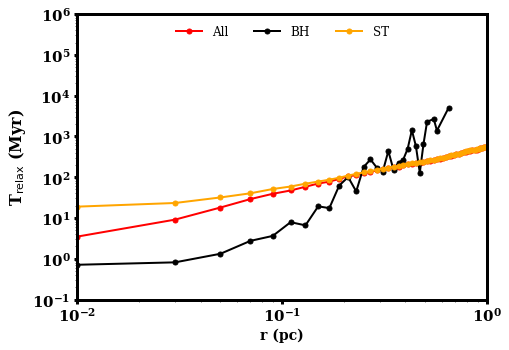}

\centering
     \includegraphics[width=0.98\hsize]{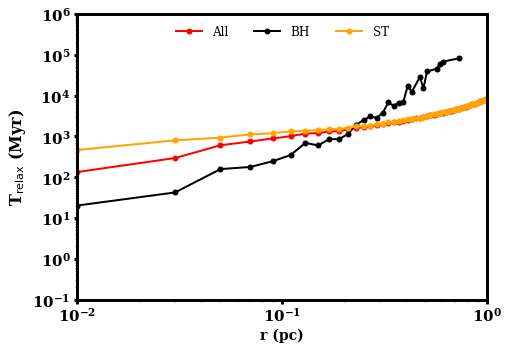}
\centering
\caption{{ The relaxation time as a function of distance from the center at initial time (\textit{top panel}) and at final time $T=34.5$~Myr (\textit{bottom panel}) for NGC205 with 1:4 mass ratio IMBHs. Relaxation time is calculated using the relation from \protect\cite{Binney2008}. The colour coding is for the three different components: BHs (black), stars (orange), and all particles (red).}}
\label{fig:205-relax-time}
\end{figure}

A BW cusp that re-establishes itself after initial erosion caused by an IMBH binary must happen at relaxation timescales. Figure~\ref{fig:205-relax-time} shows the relaxation time estimates for various components at $T = 0$~Myr (\textit{top panel}) and $T=34$~Myr (\textit{bottom panel}) for the model with the BH component. We observe that as the radius approaches 0.01~pc, the relaxation time for the BH component is initially on the order of 1~Myr, increasing to around 10~Myr in the later stages of the IMBH binary's evolution, after it has eroded the central cusp. These timescales are consistent with the formation of a BW cusp around 10~Myr, as demonstrated in the Fig.~\ref{fig:power-index-205}.

Now we discuss how the presence of a BW cusp for the BH component impacts the IMBH binary evolution, specifically focusing on the hardening rates. The hardening rate of an IMBH binary is expected to be related to properties of galactic nuclei through,
\begin{equation}
    s = GH \frac{\rho}{\sigma}, 
\label{eq:srho}
\end{equation}
where $\rho$ and $\sigma$ are density and velocity dispersion, and both these quantities are typically measured at the binary's sphere of influence \citep{seskha+15}. $H$ is a dimensionless hardening parameter, and its value for the full loss cone regime is $\approx 14$ as suggested by the scattering experiments \citep{quinlan+96,sesana+10}.     

\begin{figure}
    \centering
  \includegraphics[width=0.98\hsize]{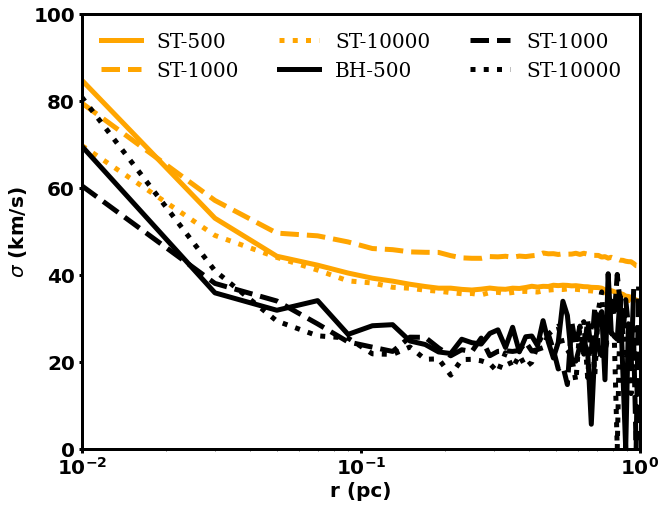}\\[1ex]
    \caption{Velocity dispersion profiles of BHs and stellar components for the NGC205 model with 1:4 mass ratio IMBHs at different times for labeled values of $1/a$.}
    \label{fig:dispersion}
\end{figure}


\begin{figure*}
\centering
\includegraphics[width=0.48\linewidth]{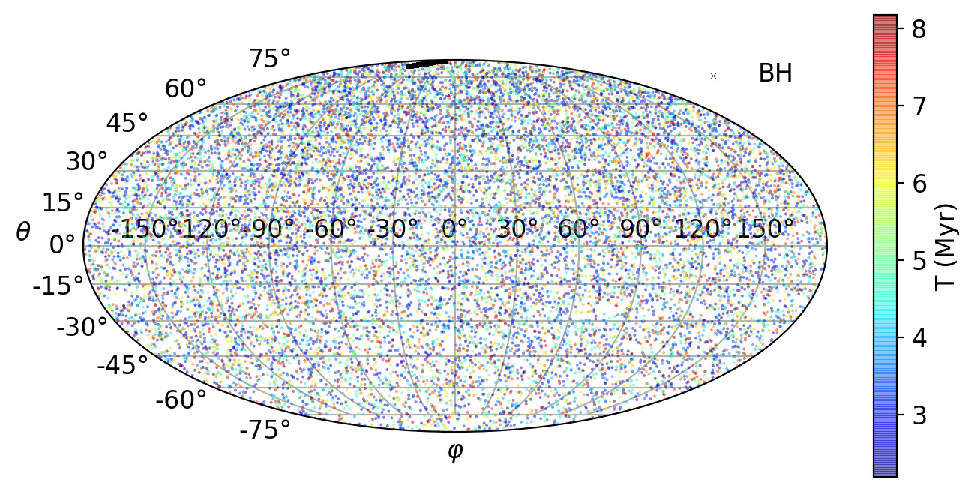}
\includegraphics[width=0.48\linewidth]{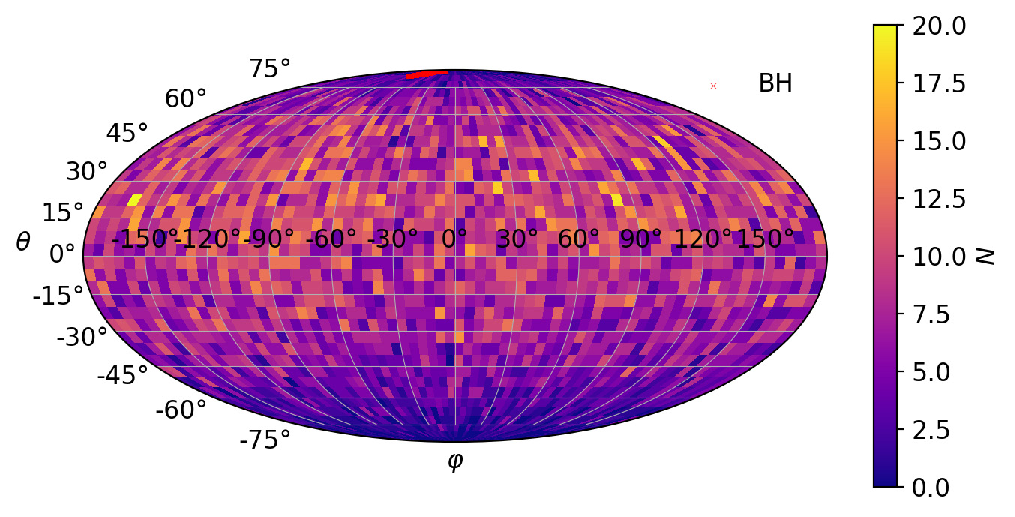}
\caption{Distribution of particles final ejection direction by the IMBH binary on the time scale $T=2$-8~Myr for NGC205 with 1:4 mass ratio. The colour code is time in the \textit{left panel} and the number of particles in each bin in the \textit{right panel}. Black (left) and red (right) crosses on the plots' top are the IMBH binary angular momentum direction. The asymmetry in positive and negative $\theta$ directions is at the level $\approx40\%$. Particles generally are ejected in the direction of the IMBH binary orbital plane for a chosen time interval.}
\label{fig:lz-stars-distr}
\end{figure*}

From Fig.~\ref{fig:205dens-stars+bh}, we see that BHs maintain a density comparable to that of the stellar population close to the center, which can lead to a similar hardening as provided by the stellar component. We also plot the velocity dispersion profiles of stellar and BH components for both of the cases as shown in Fig.~\ref{fig:dispersion}. Clearly, the BH component has lower velocity dispersion compared to the stellar component for the same values of $1/a$. The lower value of the velocity dispersion boosts the hardening rate contribution by the BHs component (see equation~\ref{eq:srho}). Thus, the higher hardening rate witnessed in the model with the BH component can be attributed to the sustained cuspy density profile of the BHs close to the center and their relatively lower velocity dispersion. 

We also plot the spatial distribution of the particles for a duration of 2-8~Myr of the run after they are scattered by the IMBH binary in a hard binary regime (see \textit{left panel} of Fig.~\ref{fig:lz-stars-distr}). Also, the \textit{right panel} of the figure shows the number of particles scattered at different angles. We see that particles are preferentially ejected in the direction of the IMBH binary plane. 



\begin{figure}
\centering
  \includegraphics[width=0.98\hsize]{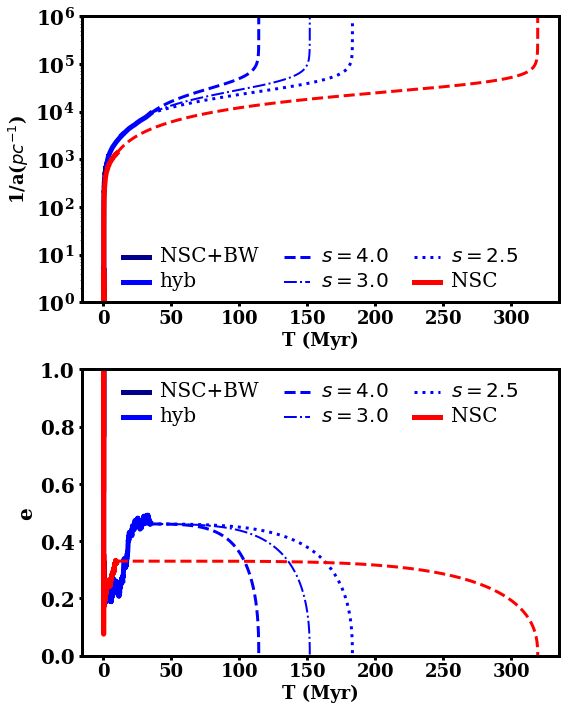}\\[1ex]
\caption{ IMBH binary parameters evolution for NGC205 model with 1:4 mass ratio IMBHs.
\textit{Top panel}: Inverse semimajor axis of IMBH binaries in our simulation (full lines) and its estimated evolution (dashed and dotted lines) taking into account energy loss through GW emission. The evolution is estimated using different values of $s$ for the NSC+BW model. \textit{Bottom panel}: Eccentricity of IMBH binaries in our simulation. The line and color scheme is the same as employed for $1/a$ plot in \textit{top panel}. 
} \label{fig:estimates205}
\end{figure}

\subsubsection{NGC205: IMBH binary (1:4) merger timescales}\label{subsec:estimates1to4}

We do not include relativistic effects such as gravitational-wave emission on the fly in our simulations. Instead, from the time we stop our runs, the IMBH binary semimajor axis and eccentricity are evolved using the prescription developed in our earlier studies \citep{khan+12a,seskha+15}. We adopt a constant value for $s$ and $e$ at the end of each simulation, whereas energy loss caused by GW emission is taken into account using the analytical expression of \citet{peters+63}. The resulting evolution of $1/a$ and $e$ is presented in \textit{top} and \textit{bottom panels} of Fig.~\ref{fig:estimates205}, respectively. For model NSC+BW, an $e$ value of $0.46$ measured at the end of our simulation, the IMBH binary achieves a merger around $115$~Myr. However, if $e$ continues to grow and reaches a moderate value of $e \approx 0.75$, then a merger can be achieved around $80$~Myr. We also use lower values of $s = 3.0$ and $2.5$ to estimate the IMBH binary merger time, resulting in IMBH coalescence in roughly 150~Myr and 200~Myr, respectively. The merger of IMBHs in the model with a stellar-only component takes place at a much longer time $T \approx 320$~Myr. This is understandable, as the higher hardening rates witnessed for the NSC+BW model translate into a shorter IMBH binary merger time, given that the eccentricity value is similar for both runs.

\subsubsection{NGC205: IMBH binary (1:40) evolution} \label{subsec:simulation1to40}

In order to investigate the effect of different mass ratios, we perform a run with a 1:40 mass ratio of IMBHs. We kept the primary IMBH mass fixed but introduced a secondary that is 40~times less massive than the primary. The hardening rates for IMBH binaries are shown as red rectangles in Fig.~\ref{fig:s0p25}. We notice roughly 3~times higher hardening rates for the NSC model only and 4~times for NSC+BW models. This is within a factor of 2 of the change in reduced mass as we moved from 1:4 to 1:40 mass ratio of IMBH binary, and inline with \citep{ber22} who found that hardening rate scales inversely with the reduced mass for a fixed density distribution of the host.  

\subsubsection{NGC205: IMBH binary (1:40) merger timescales} \label{subsec:estimates1to40}

\begin{figure}
\centering
  \includegraphics[width=0.98\hsize]{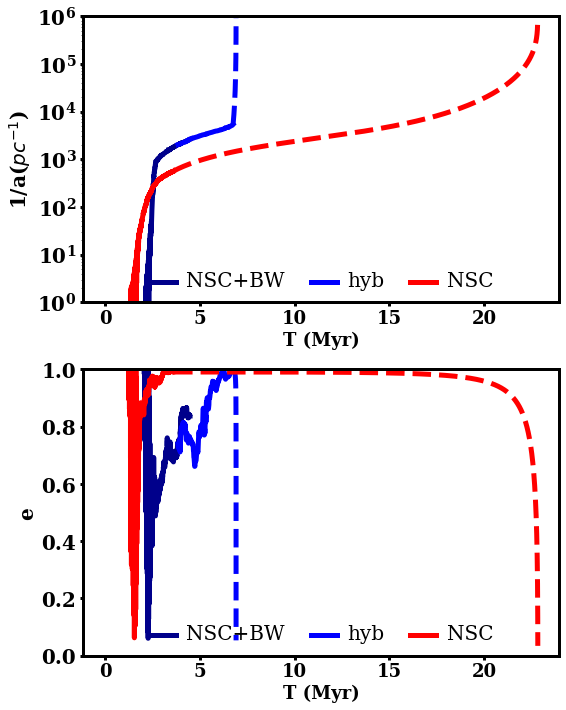}\\[1ex]
\caption{
Top panel: IMBH binaries inverse semimajor axis in our simulation (full lines), and its estimated evolution (dashed lines) for NGC205 models with 1:40 mass ratio IMBHs. Bottom panel: Eccentricity evolution of IMBH binaries. 
} \label{fig:205estimates1to40}
\end{figure}

Figure~\ref{fig:205estimates1to40} shows the complete evolution of the IMBH binaries, including the estimated evolution taking into account the GW emission. For both cases, with and without BH components, we notice rapid eccentricity growth after binary formation in the three-body scattering phase. In both cases, the eccentricity $e$ approaches unity, leading to a rapid decay of the IMBH binary orbits through enhanced gravitational wave emission. The eccentricity reaching a very high value close to unity for high mass ratios has previously been reported in $N$-body simulations of SMBH binaries both in isolated galaxy models \citep{Iwasawa+11}, and also in galaxy mergers \citep{Khan+15}.

The coalescence times of IMBHs are $7$ and $23$~Myr for models with and without the BH component, respectively, significantly shortened by a very high value of $e$.   

\subsection{IMBH Binary evolution - NGC404} \label{subsec:404}

\begin{figure*}
    \centering
     \includegraphics[width=0.98\hsize]{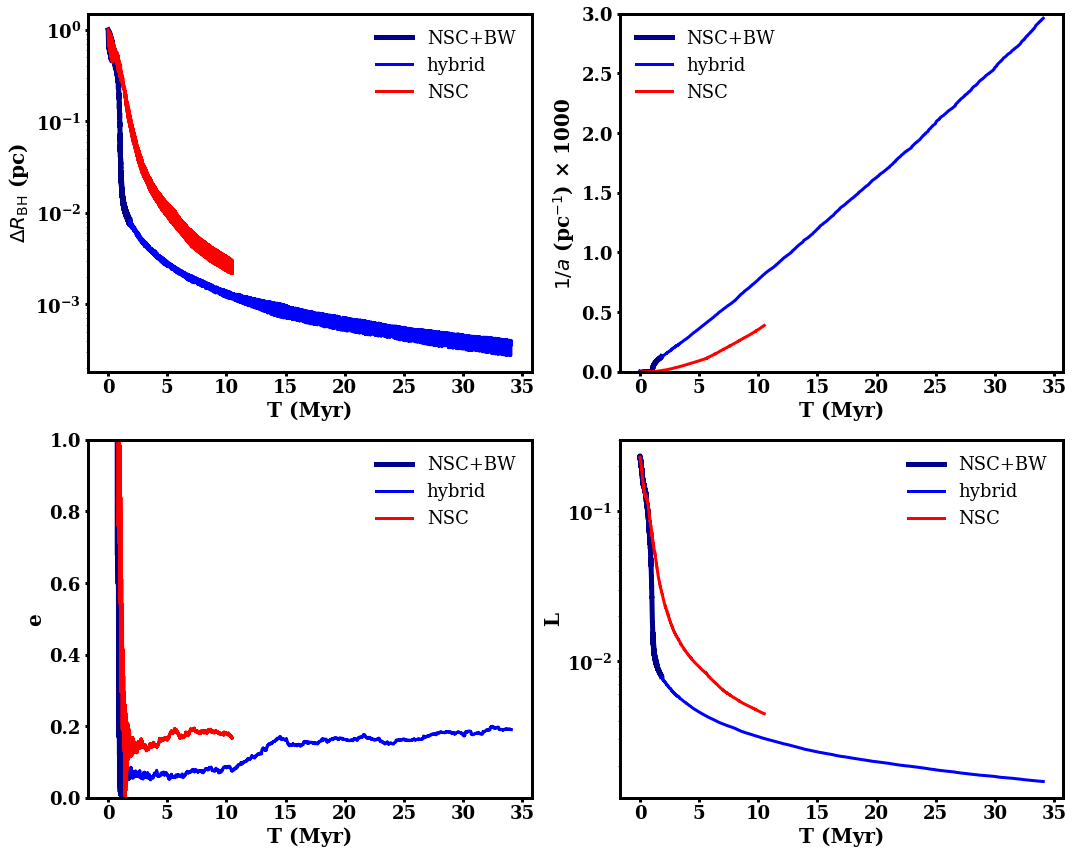}
    \caption{IMBH pair and binary evolution in NGC404 models (with and without a central BH cusp). \textit{Top left}: IMBHs orbit relative separation ($\Delta R_{\rm BH}$). \textit{Top right}: IMBHs orbit inverse semimajor axis ($1/a$). \textit{Bottom left}: IMBH binaries eccentricity ($e$). \textit{Bottom right}: Total angular momentum ($L$) of IMBH binaries in units of $5.89 \times 10^8$ $M{\odot}$~pc~km/s.}
    \label{fig:param404}
\end{figure*}

The evolution of the IMBH binary for NGC404 NSC models is presented in Fig.~\ref{fig:param404}. We observe a slower shrinking of the IMBH binary separation (\textit{top left panel}) in the absence of a  BH component, compared to that where a BH component is present. A similar trend is also observed in the evolution of $1/a$ (\textit{top right panel}) and angular momentum (\textit{bottom right panel}), where the IMBH binary with BH component evolves more rapidly. This behavior can be explained by examining the density profiles of the stellar and BH populations in the \textit{top panel} of Fig.~\ref{fig:stab404}. The density is high and has a steeper slope in the case of the BH component, whereas it is lower with a much shallower slope in the stellar component. As discussed earlier, the evolution of the IMBH binary strongly depends on the density profile of the background mass distribution in both the dynamical friction \citep{just11} and the three-body hardening phase. IMBH binary eccentricity remains low for the duration of our runs.

\begin{figure}

 \centering
     \includegraphics[width=0.98\hsize]{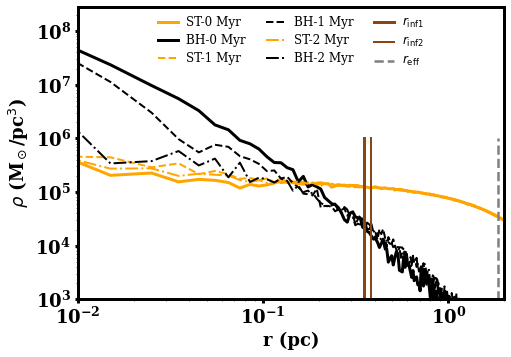}
\centering
     \includegraphics[width=1.02\hsize]{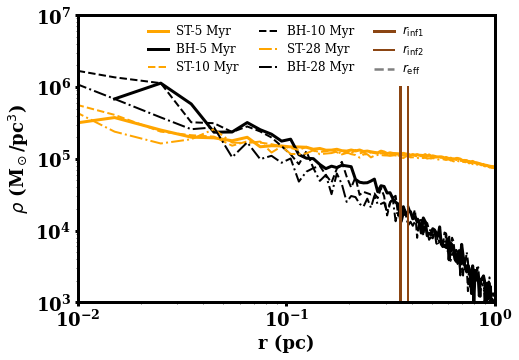}
  
\caption{ Density profile for NGC404 model.
\textit{Top panel}: Density profile of stellar and BH components in NSC+BW before and after a hard binary forms.
\textit{Bottom panel}: Time evolution of stellar and BH component density profile in the late phase of IMBH binary evolution.} \label{fig:dens404-stars+bh}
\end{figure}

Again, as in earlier cases, we estimated IMBH binary hardening rates $s$ for both runs by fitting a straight line at various times (Fig.~\ref{fig:s0p25}). Again, we witness a higher value of $s$ for IMBH binary evolving in NSC+BW model (filled blue circles with full line). The hardening rates remain steady and start to increase towards the end of our run. This trend is consistent with what we witnessed for the NGC205 model with a BH component. The value of $s$ remains almost constant for the NSC model without the BH component.

We analysed the evolution of the density profile for the NSC+BW model as shown in Fig.~\ref{fig:dens404-stars+bh}. The \textit{top panel} of the figure showcases the density evolution of both stellar and BH components as the binary forms and transitions to a hard binary phase ($T \approx 2$~Myr). We notice that the BH component got significantly eroded while the stellar profile changes relatively less strongly. In the \textit{bottom panel} of the figure, which presents the evolution in a hard binary regime, we only see a very slight decrease in BH stellar density profiles, whereas again the stellar component retains its profile. In the hard binary regime, IMBH binary's semimajor axis shrank by an order of magnitude. However, we do not witness a noticeable change in both the stellar and BH profiles close to the center. This suggests that energy deposited by the IMBH binary is distributed at much larger spatial scales. 

\begin{figure}
\centering
     \includegraphics[width=0.98\hsize]{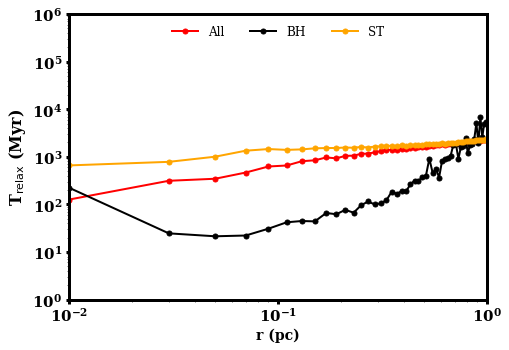}

\centering
     \includegraphics[width=0.98\hsize]{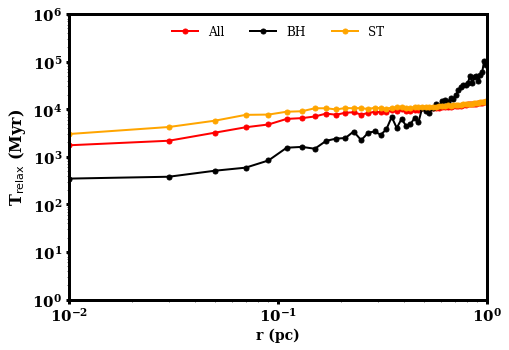}
\centering
\caption{ { The relaxation time at various distances from the center at initial time (\textit{top panel}) and at final time $T=27.5$~Myr (\textit{bottom panel}) for NGC404. Relaxation time is calculated using \protect\cite{Binney2008}. The colour coding is the three different components: BHs (black), stars (orange), and all particles (red).}}
\label{fig:relax-time-ngc404}
\end{figure}

Again, for NGC404, we calculate relaxation times (see Fig.~\ref{fig:relax-time-ngc404}) for both the stellar and BH components at the start of the simulation (\textit{top panel}) and later in the hard binary regime (\textit{bottom panel}). Initially, within the inner 0.1~pc (approximately the influence radius of the IMBH), the relaxation time is around 10~Myr for the BH component and a few hundred Myr for the stellar component. As the IMBH binary erodes the central cusp, these relaxation times increase by an order of magnitude.

\begin{figure}
\centering
     \includegraphics[width=0.98\hsize]{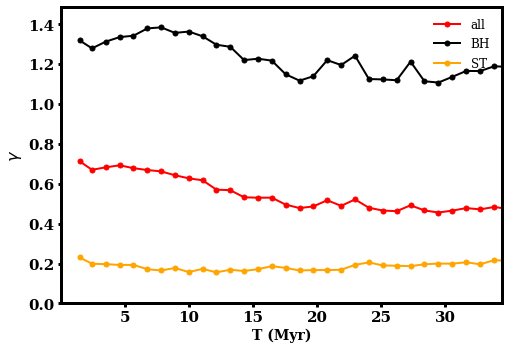}
\centering
\caption{
{Evolution of the inner logarithmic density slope parameter $\gamma$ for NGC404 NSC+BW run. The parameter $\gamma$ was calculated by fitting and averaging over the nine snapshots in a timespan of $\approx 0.1$~Myr. The errors of the averaging are less than one per cent. The colour coding represents three different components: BHs (black), stars (orange), and all particles (red). }
} \label{fig:404index}
\end{figure}

Because the relaxation time is longer than the duration of our run, the BH component does not achieve the BW cusp level as seen from Fig.~\ref{fig:404index}. The BH component has the steepest slope in density between -1 and -1.5, whereas the stellar component has a much shallower slope of -0.25.  

\begin{figure}
\centering
     \includegraphics[width=0.98\hsize]{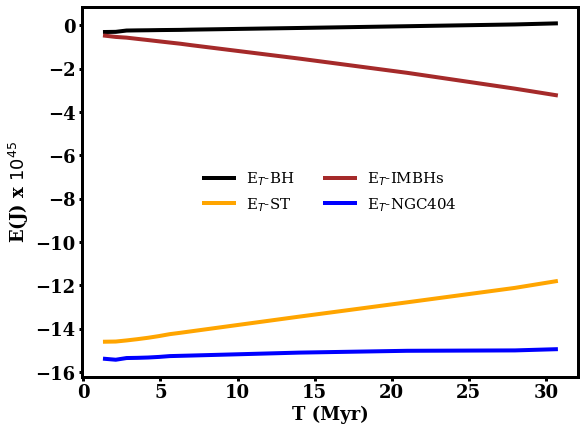}
\caption{
Total energy of the whole system (blue), stellar component (orange), BH component (black), and IMBH binary (brown) for the whole duration of our simulation for the NGC404 NSC+BW run. 
} \label{fig:404energy}
\end{figure}

We investigate the energy changes of individual species and those of the entire system and present them in Fig.~\ref{fig:404energy}. We start our discussion with the energy change of the IMBH binary (olive line). We notice that the total energy change of the IMBH binary from a hard binary until we stop our run is roughly 0.02 in our model units. This energy must be deposited into the stellar and BH components to maintain energy conservation. If we look at how the energy of the BH component evolves (black line), we notice that the gain is only a fraction of the total change experienced by the IMBH binary. Even if the BH component deposits all its binding energy in the IMBH binary, it would not significantly change the IMBH binary energy because one can witness from Fig.~\ref{fig:404energy} that the BH component only carries a tiny fraction of the total energy of the system. We notice that the total energy of the BH component becomes positive, but it does not mean that the BH component has become unbound from the system, as we can still see a strong cusp of BHs around the IMBH binary in Figs.~\ref{fig:dens404-stars+bh} and \ref{fig:404index}. This net total positive energy of the BH component should be caused by some strong encounters with the IMBH, resulting in a large gain in kinetic energy. However, the total energy gained by the stellar component (brown line) is similar to that of energy loss by the IMBH binary. This suggests that most of the energy lost by the IMBH binary is deposited into the stellar component. This raises the question of why we do not see a significant impact on stellar density profiles in the vicinity of the IMBH binary. Also, for the case where no BH component is present, we notice significantly smaller hardening rates. So indeed BH component contributes to the hardening of IMBH binary as much as the stellar component, if not more, but why does not the energy budget of the individual components reflect the same picture? 

Our explanation for this discrepancy is that although BHs suck most of the IMBH binary energy through three-body encounters, they sink back towards the center through collisional mass segregation with the stellar component on a relatively short timescales, as they are 10 times more massive than the stellar particles. In the process, they deposit the energy in the stellar component. The profiles for both the stellar and BH components remain more or less the same; for BHs, because they constantly move in and due to mass segregation and three-body scattering with the IMBH binary, and for stars, because they can exchange energy from the BH component at larger distances. Additionally, it is reassuring to witness that the total energy of the whole system remains constant during the whole run. A related mechanism, operating on a different scale during star-cluster assembly, involves low-mass IMBH ($\sim100$ -- $1000 M_\odot$) triple system in which one member is ejected and later returns to the center after losing energy, a process highlighted as an important channel for IMBH growth through mergers in hierarchical triple black-hole systems \citep{rant24,liu2024,sou2025}.

\begin{figure}
\centerline{
  \resizebox{0.98\hsize}{!}{\includegraphics[angle=0]{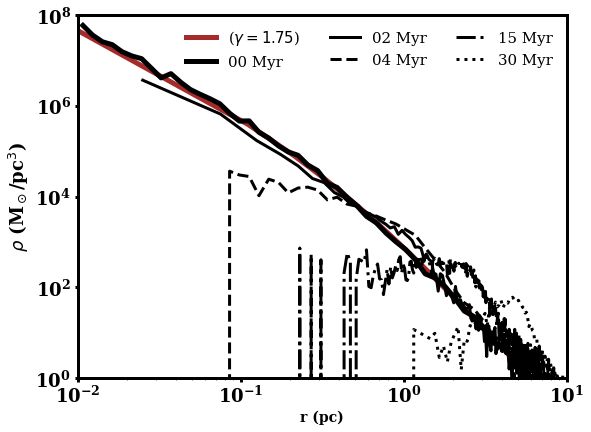}}
  }
\caption{
Density changes as a function of time (in Myr) induced by IMBH binary in BH component in the absence of a stellar mass distribution (realized as an external potential) for the NGC404 NSC+BW run.
} \label{fig:404rhobhs}
\end{figure}

To test our hypothesis, we ran an additional run for the NGC404 model, this time with the BH component only. To make our model consistent with the previous model, we introduce a stellar component as an external potential so that its gravitational potential is there, but it will not undergo energy exchanges with the IMBH binary, as well as with the BH component. Figure~\ref{fig:404rhobhs} shows how the density of the BH component evolves in case there is no stellar component to exchange energy with. We see that over time, IMBH binary scours the core, and by the end of the simulation, at a time corresponding to 30~Myr, the cusp is totally erased inside the inner 1~pc. This strengthens our understanding that energy exchange between stellar and BH components, together with mass segregation, is key to maintaining the BH cusp around the IMBH binary, resulting in efficient hardening of the binary. 
 
\subsubsection{NGC404: IMBH binary (1:4) merger timescales} \label{sub:404merger}

The eccentricity of the binary remains very low in both cases, with and without a BW cusp. The hardening rates (in model units) for the binary are $0.76$ with the BH component and $0.46$ without the BH component. IMBH binary hardening rate in the model with BH component increases with time and reaches a value of 0.98 at the time we stop the run (Fig.~\ref{fig:s0p25}).

\begin{figure}
\centering
  \includegraphics[width=0.98\hsize]{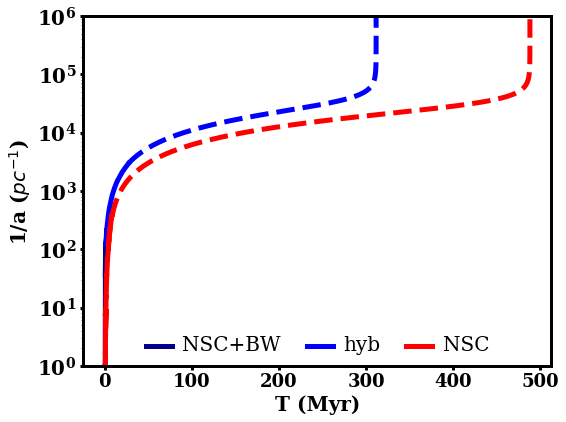}\\[1ex]
  
\caption{
Inverse semimajor axis of IMBH binaries in our simulation (full lines) and its estimated evolution (dashed lines) for the NGC404 NSC+BW model. 
} \label{fig:404estimates1to4}
\end{figure}

We evolve the binary in both scenarios using these hardening rates and accounting for gravitational wave (GW) emission. The resulting evolution is depicted in Fig.~\ref{fig:404estimates1to4}. We observe that the merger time for IMBHs in the stellar component model is almost twice as long as the model that includes a BH component. Because the eccentricities are very similar, this difference is directly related to the variation in hardening rates.

\section{Summary and Conclusion} \label{sec:summary}

We perform high-resolution $N$-body simulations of an Intermediate Mass Black Hole (IMBH) pair in nuclear star clusters (NSC's). The structural and kinematic parameters are motivated by the observations of NSCs in nearby dwarf galaxies, namely NGC205 and NGC404. For the first time in numerical simulations of IMBH dynamics in NSCs, we achieve a mass resolution of 1~M$_{\odot}$ for the stellar component. We also include a dark stellar mass BH component with a Bahcall-Wolf (BW) cusp inside the influence radius of the central IMBH. The black hole component has a mass of one per cent of that of the NSC. Well inside the influence radius, stellar BHs dominate in the mass density, and as we approach the influence radius, the stellar component takes over (see Figs.~\ref{fig:stab205} and \ref{fig:stab404}). The BH particles are realized with 10~M$_{\odot}$~particles. To make a comparison and better understand the impact of the BH component, we also study the evolution of the IMBH pair for each model where the BH component is absent. We introduce the secondary IMBH with varying masses on moderately elliptical orbits ($e \approx 0.6$) at an initial separation of 1~pc. We assume that IMBHs can be delivered to the center of dwarf galaxies either by inspiralling star clusters and/or mergers between dwarf galaxies. Dense star clusters are known to form an IMBH quite quickly, see the paragraph and citations at the end of Sect.~\ref{orbits}. In the following, we summarize key findings of the study:

\begin{itemize}
    \item IMBHs form a bound system in a few Myrs mainly because the secondary IMBH is already inside the effective radius of the NSC, where high stellar and BH densities result in shorter dynamical friction time.

    \item An IMBH binary quickly disrupts the stellar and BH cusps as it enters a hard binary phase, resulting in a density drop of both the stellar and BH components by an order of magnitude in the center. The BH densities dominate very close to the IMBH binary ($r \approx 0.01$~pc), but as we move out to the influence radius, stellar densities start to take over (see Figs.~\ref{fig:205dens-stars+bh} and \ref{fig:dens404-stars+bh}). This is true for both the NGC205 and NGC404 cases.

    \item IMBH binaries form and maintain low to moderate eccentricities for comparable masses of IMBHs for both cases. However, for the more extreme mass ratio (1:40) in NGC205, the IMBH binaries achieve very high eccentricity that approaches unity in the hard binary phase. 
    

    \item The presence of a BH component significantly enhances hardening rates compared to the case where a BH component is absent. Also, the trend of the hardening rate as a function of time in the hard binary regime is different. For models with a BH component, hardening rate increases with time, and for models without a BH component, it decreases more in line with earlier studies of SMBH binaries in galactic nuclei.  

    \item Due to a very short relaxation time, especially in NGC205, the BH component reestablishes a BW cusp, which is initially eroded by the IMBH binary. The cusp stays there for the whole duration of our run, resulting in an efficient energy extraction from the IMBH binary's orbit. 

    \item Although BHs are only one per cent of the stellar component, and even if they lose all of their binding energy, it would account for a tiny fraction of IMBH binary's orbital energy. However, we discover an interesting mechanism in which they carry away energy from the binary, exchange it with the stellar component, and segregate back towards the center, thus generating an efficient cycle of energy loss for the IMBH binary.

    \item Due to higher hardening rates, IMBH binaries evolving in models with a BH component attain mergers in timescales that are roughly twice as short as their counterparts that do not include a BH component.

    \item For high mass ratios, IMBH binaries achieve merger in very short time scales close to 10~Myr, aided by very high eccentricity that approaches unity. The timescales are short for both cases, with and without a BH component, caused by a very high value of eccentricity. This may have important consequences for the growth of IMBHs in NSCs, as very short merger timescales provide a very viable channel for the growth mechanism through mergers with compact objects. 
    \item The shorter merger time (by a factor of 15-20) in cases where the secondary IMBH is much lighter than the primary (1:40) can potentially boost Intermediate Mass Ratio Inspirals (IMRIs). Here, again presence of a BH component significantly shortens the binary merger time. 
    \item The IMBH binary mergers found in our simulations should be observable at the upper frequency end of future space based instruments (LISA, Tianqin and Taiji) as well as future ground based Einstein Telescope (ET) and Cosmic Explorer (CE), see citations in Sect.~\ref{sec-intro}. In the best case the combination of all these instruments will deliver a census of IMBH across a large redshift range, which makes it important to start early enough with detailed numerical modeling and accurate estimates of merger timescales \citep{KHB2025}.
\end{itemize}


Our models adopt single-mass stellar particles and a single-mass black hole component. This allows us to study the dynamical interaction between the IMBH binary and the surrounding subsystems, but does not capture the full complexity of a realistic stellar initial mass function. A high-mass tail of stellar-mass black holes could lead to formation of triple systems and potentially affect the orbital decay. Additionally, mass segregation of BHs scattered by an IMBH binary is sensitive to the individual BH masses: more massive BHs reach the center faster and can accelerate energy transfer from the IMBH binary, whereas less massive BHs lead to longer decay times and less efficient energy extraction. Whether these effects offset each other or one dominates is an interesting question, and we will investigate the impact of a realistic BH mass function on IMBH binary evolution in a forthcoming study.

Although $\phi-$GPU is capable of modeling IMBHs dynamics in the relativistic regime with PN terms up to 3.5, here we do not include this effect primarily because these terms become relevant only once the IMBH binary reaches semi-major axes $\sim 10^{-5}$ pc, much smaller than a typical value of $\sim 10^{-4}$ pc achieved in our simulations. Our work focuses on the Newtonian hardening phase; and the subsequent evolution of IMBH binaries is treated semi-analytically.

NSCs in dwarf galaxies are expected to experience episodic gas inflows which boost star formation and IMBH growth through gas accretion \citep{Partmann2025}. Hydrodynamics would introduce additional drag and torques, accelerating the merger of IMBHs in the binary \citep{duffell20,bortolas21,franchini22,siwek23}.
In both systems considered here, the nuclear star clusters are dynamically dominated by stars, with gas contributing only a minor fraction (a few percent) of the central mass \citep{De2012,Davis2020}. In this regime, stellar dynamics is expected to be the primary driver of IMBH binary evolution.

We model IMBH binary dynamics in an isolated NSC environment. The outer dwarf-galaxy potential is ignored because our modeled NSCs are much denser and dynamically dominant in the central parsecs. In our previous study \citet{Khan_Holley-Bockelmann2021}, we demonstrated that for NSC/IMBH mass ratios greater than 10, which is the case for both NGC205 and NGC404, the contribution of encounters from the surrounding galaxy is not significant.

Overall, while these simplifications limit the completeness of the physical processes involved in IMBHs dynamics in NSCs, they do not affect the mechanism identified in this work--namely, the enhanced hardening and shortened merger times of the IMBH binary driven by the centrally concentrated cusp of stellar-mass black holes. Future simulations that incorporate a full mass spectrum and relativistic effects will enable a comprehensive examination of IMBH binary evolution in more realistic nuclear star cluster environments.


%





\begin{acknowledgements}

The authors gratefully acknowledge the Gauss Centre for Supercomputing e.V. (www.gauss-centre.eu) for funding this project by providing computing time on the GCS Supercomputer JUWELS \citep{JUWELS} at J\"ulich Supercomputing Centre (JSC). We acknowledge support from High Performance Computing resources at New York University Abu Dhabi. This material is based upon work supported by Tamkeen under the NYU Abu Dhabi Research Institute grant CASS. Rainer Spurzem is grateful for support by the German Science Foundation (DFG), grant No. Sp 345/24-1, and acknowledges NAOC (National Astronomical Observatories of Chinese Academy of Sciences, Beijing) International Cooperation Office for its support in 2023, 2024, and 2025, and the support by the National Science Foundation of China (NSFC) under grant No. 12473017. Peter Berczik and Margarita Sobolenko thank the support from the special program of the Polish Academy of Sciences and the U.S. National Academy of Sciences under the Long-term program to support Ukrainian research teams, grant No.~PAN.BFB.S.BWZ.329.022.2023. 

\end{acknowledgements}

\bibliographystyle{mnras}  
\bibliography{ms}   

@ARTICLE{Molina21,
       author = {{Molina}, Mallory and {Reines}, Amy E. and {Latimer}, Lilikoi J. and {Baldassare}, Vivienne and {Salehirad}, Sheyda},
        title = "{A Sample of Massive Black Holes in Dwarf Galaxies Detected via [Fe X] Coronal Line Emission: Active Galactic Nuclei and/or Tidal Disruption Events}",
      journal = {\apj},
         year = 2021,
        month = dec,
       volume = {922},
       number = {2},
          eid = {155},
        pages = {155},
          doi = {10.3847/1538-4357/ac1ffa},
archivePrefix = {arXiv},
       eprint = {2108.09307},
 primaryClass = {astro-ph.GA},
       adsurl = {https://ui.adsabs.harvard.edu/abs/2021ApJ...922..155M}
}

@BOOK{Binney2008,
       author = {{Binney}, James and {Tremaine}, Scott},
        title = "{Galactic Dynamics: Second Edition}",
         year = 2008,
    publisher = {Princeton University Press, Princeton, NJ},
       adsurl = {https://ui.adsabs.harvard.edu/abs/2008gady.book.....B}
}

@ARTICLE{Bahcall1976,
       author = {{Bahcall}, J.~N. and {Wolf}, R.~A.},
        title = "{Star distribution around a massive black hole in a globular cluster.}",
      journal = {\apj},
         year = 1976,
        month = oct,
       volume = {209},
        pages = {214-232},
          doi = {10.1086/154711},
       adsurl = {https://ui.adsabs.harvard.edu/abs/1976ApJ...209..214B}
}

@ARTICLE{vasi19,
       author = {{Vasiliev}, Eugene},
        title = "{AGAMA: action-based galaxy modelling architecture}",
      journal = {\mnras},
         year = 2019,
        month = jan,
       volume = {482},
       number = {2},
        pages = {1525-1544},
          doi = {10.1093/mnras/sty2672},
archivePrefix = {arXiv},
       eprint = {1802.08239},
 primaryClass = {astro-ph.GA},
       adsurl = {https://ui.adsabs.harvard.edu/abs/2019MNRAS.482.1525V}
}

@INPROCEEDINGS{Berczik2013,
  author={{Berczik}, P. and {Spurzem}, R. and {Zhong}, S. and {Wang}, L. and {Nitadori}, K. and {Hamada}, T. and {Veles}, A.},
  title="{Up to 700k GPU cores, Kepler, and the Exascale future for simulations of star clusters around black holes}",
  booktitle = {Procs. of 28th Intl. Supercomputing Conf. ISC 2013, Leipzig, Germany, June 16-20, 2013.},
  year = 2013,
  series = {Lecture Notes in Computer Science},
  volume = 7905,
  editor = { {Kunkel}, Julian M. and {Ludwig}, Thomas and {Meuer}, Hans (Eds.)},
  month = jun,
  pages = {13-25},
  publisher = {Springer Vlg.}
}

@ARTICLE{khan+12a,
   author = {{Khan}, F.~M. and {Preto}, M. and {Berczik}, P. and {Berentzen}, I. and 
	{Just}, A. and {Spurzem}, R.},
    title = "{Mergers of Unequal-mass Galaxies: Supermassive Black Hole Binary Evolution and Structure of Merger Remnants}",
  journal = {\apj},
archivePrefix = "arXiv",
   eprint = {1202.2124},
     year = 2012,
    month = apr,
   volume = 749,
      eid = {147},
    pages = {147},
      doi = {10.1088/0004-637X/749/2/147},
   adsurl = {http://adsabs.harvard.edu/abs/2012ApJ...749..147K}
}

@ARTICLE{graham+15,
   author = {{Graham}, M.~J. and {Djorgovski}, S.~G. and {Stern}, D. and 
	{Glikman}, E. and {Drake}, A.~J. and {Mahabal}, A.~A. and {Donalek}, C. and 
	{Larson}, S. and {Christensen}, E.},
    title = "{A possible close supermassive black-hole binary in a quasar with optical periodicity}",
  journal = {\nat},
archivePrefix = "arXiv",
   eprint = {1501.01375},
     year = 2015,
    month = feb,
   volume = 518,
    pages = {74-76},
      doi = {10.1038/nature14143},
   adsurl = {http://adsabs.harvard.edu/abs/2015Natur.518...74G}
}

@ARTICLE{khan+11,
   author = {{Khan}, F.~M. and {Just}, A. and {Merritt}, D.},
    title = "{Efficient Merger of Binary Supermassive Black Holes in Merging Galaxies}",
  journal = {\apj},
archivePrefix = "arXiv",
   eprint = {1103.0272},
     year = 2011,
    month = may,
   volume = 732,
      eid = {89},
    pages = {89},
      doi = {10.1088/0004-637X/732/2/89},
   adsurl = {http://adsabs.harvard.edu/abs/2011ApJ...732...89K}
}

@ARTICLE{peters+63,
   author = {{Peters}, P.~C. and {Mathews}, J.},
    title = "{Gravitational Radiation from Point Masses in a Keplerian Orbit}",
  journal = {Physical Review},
     year = 1963,
    month = jul,
   volume = 131,
    pages = {435-440},
      doi = {10.1103/PhysRev.131.435},
   adsurl = {http://adsabs.harvard.edu/abs/1963PhRv..131..435P}
}

@ARTICLE{quinlan+96,
   author = {{Quinlan}, G.~D.},
    title = "{The dynamical evolution of massive black hole binaries I. Hardening in a fixed stellar background}",
  journal = {\na},
   eprint = {astro-ph/9601092},
     year = 1996,
    month = jul,
   volume = 1,
    pages = {35-56},
      doi = {10.1016/S1384-1076(96)00003-6},
   adsurl = {http://adsabs.harvard.edu/abs/1996NewA....1...35Q}
}

@ARTICLE{begelman+80,
   author = {{Begelman}, M.~C. and {Blandford}, R.~D. and {Rees}, M.~J.},
    title = "{Massive black hole binaries in active galactic nuclei}",
  journal = {\nat},
     year = 1980,
    month = sep,
   volume = 287,
    pages = {307-309},
      doi = {10.1038/287307a0},
   adsurl = {http://adsabs.harvard.edu/abs/1980Natur.287..307B}
}

@ARTICLE{berczik+06,
   author = {{Berczik}, P. and {Merritt}, D. and {Spurzem}, R. and {Bischof}, H.-P.
	},
    title = "{Efficient Merger of Binary Supermassive Black Holes in Nonaxisymmetric Galaxies}",
  journal = {\apjl},
   eprint = {astro-ph/0601698},
     year = 2006,
    month = may,
   volume = 642,
    pages = {L21-L24},
      doi = {10.1086/504426},
   adsurl = {http://adsabs.harvard.edu/abs/2006ApJ...642L..21B}
}

@INPROCEEDINGS{berczik+11,
   author = {{Berczik}, P. and {Nitadori}, K. and {Zhong}, S. and {Spurzem}, R. and 
	{Hamada}, T. and {Wang}, X. and {Berentzen}, I. and {Veles}, A. and 
	{Ge}, W.},
    title = "{High performance massively parallel direct N-body simulations on large GPU clusters.}",
booktitle = {International conference on High Performance Computing, Kyiv, Ukraine, October 8-10, 2011., p. 8-18},
     year = 2011,
    month = oct,
    pages = {8-18},
   adsurl = {http://adsabs.harvard.edu/abs/2011hpc..conf....8B}
}

@ARTICLE{seskha+15,
   author = {{Sesana}, A. and {Khan}, F.~M.},
    title = "{Scattering experiments meet N-body - I. A practical recipe for the evolution of massive black hole binaries in stellar environments}",
  journal = {\mnras},
archivePrefix = "arXiv",
   eprint = {1505.02062},
     year = 2015,
    month = nov,
   volume = 454,
    pages = {L66-L70},
      doi = {10.1093/mnrasl/slv131},
   adsurl = {http://adsabs.harvard.edu/abs/2015MNRAS.454L..66S}
}

@ARTICLE{sesana+10,
   author = {{Sesana}, A.},
    title = "{Self Consistent Model for the Evolution of Eccentric Massive Black Hole Binaries in Stellar Environments: Implications for Gravitational Wave Observations}",
  journal = {\apj},
archivePrefix = "arXiv",
   eprint = {1006.0730},
 primaryClass = "astro-ph.CO",
     year = 2010,
    month = aug,
   volume = 719,
    pages = {851-864},
      doi = {10.1088/0004-637X/719/1/851},
   adsurl = {http://adsabs.harvard.edu/abs/2010ApJ...719..851S}
}

@ARTICLE{Kormendy+13,
   author = {{Kormendy}, J. and {Ho}, L.~C.},
    title = "{Coevolution (Or Not) of Supermassive Black Holes and Host Galaxies}",
  journal = {\araa},
archivePrefix = "arXiv",
   eprint = {1304.7762},
     year = 2013,
    month = aug,
   volume = 51,
    pages = {511-653},
      doi = {10.1146/annurev-astro-082708-101811},
   adsurl = {http://adsabs.harvard.edu/abs/2013ARA%26A..51..511K}
}

@ARTICLE{kha16,
       author = {{Khan}, Fazeel Mahmood and {Fiacconi}, Davide and {Mayer}, Lucio and
         {Berczik}, Peter and {Just}, Andreas},
        title = "{Swift Coalescence of Supermassive Black Holes in Cosmological Mergers of Massive Galaxies}",
      journal = {\apj},
         year = "2016",
        month = "Sep",
       volume = {828},
       number = {2},
          eid = {73},
        pages = {73},
          doi = {10.3847/0004-637X/828/2/73},
archivePrefix = {arXiv},
       eprint = {1604.00015},
 primaryClass = {astro-ph.GA},
       adsurl = {https://ui.adsabs.harvard.edu/abs/2016ApJ...828...73K}
}

@ARTICLE{Gualandris+12,
   author = {{Gualandris}, A. and {Merritt}, D.},
    title = "{Long-term Evolution of Massive Black Hole Binaries. IV. Mergers of Galaxies with Collisionally Relaxed Nuclei}",
  journal = {\apj},
archivePrefix = "arXiv",
   eprint = {1107.4095},
     year = 2012,
    month = jan,
   volume = 744,
      eid = {74},
    pages = {74},
      doi = {10.1088/0004-637X/744/1/74},
   adsurl = {http://adsabs.harvard.edu/abs/2012ApJ...744...74G}
}

@ARTICLE{Khan+15,
   author = {{Khan}, F.~M. and {Holley-Bockelmann}, K. and {Berczik}, P.},
    title = "{Ultramassive Black Hole Coalescence}",
  journal = {\apj},
archivePrefix = "arXiv",
   eprint = {1405.6425},
     year = 2015,
    month = jan,
   volume = 798,
      eid = {103},
    pages = {103},
      doi = {10.1088/0004-637X/798/2/103},
   adsurl = {http://adsabs.harvard.edu/abs/2015ApJ...798..103K}
}

@ARTICLE{Iwasawa+11,
   author = {{Iwasawa}, M. and {An}, S. and {Matsubayashi}, T. and {Funato}, Y. and 
	{Makino}, J.},
    title = "{Eccentric Evolution of Supermassive Black Hole Binaries}",
  journal = {\apjl},
archivePrefix = "arXiv",
   eprint = {1011.4017},
     year = 2011,
    month = apr,
   volume = 731,
      eid = {L9},
    pages = {L9},
      doi = {10.1088/2041-8205/731/1/L9},
   adsurl = {http://adsabs.harvard.edu/abs/2011ApJ...731L...9I}
}

@ARTICLE{Antonini+12,
   author = {{Antonini}, F. and {Merritt}, D.},
    title = "{Dynamical Friction around Supermassive Black Holes}",
  journal = {\apj},
archivePrefix = "arXiv",
   eprint = {1108.1163},
     year = 2012,
    month = jan,
   volume = 745,
      eid = {83},
    pages = {83},
      doi = {10.1088/0004-637X/745/1/83},
   adsurl = {http://adsabs.harvard.edu/abs/2012ApJ...745...83A}
}

@ARTICLE{deh93,
   author = {{Dehnen}, W.},
    title = "{A Family of Potential-Density Pairs for Spherical Galaxies and Bulges}",
  journal = {\mnras},
     year = 1993,
    month = nov,
   volume = 265,
    pages = {250},
      doi = {10.1093/mnras/265.1.250},
   adsurl = {http://adsabs.harvard.edu/abs/1993MNRAS.265..250D}
}

@ARTICLE{just11,
   author = {{Just}, A. and {Khan}, F.~M. and {Berczik}, P. and {Ernst}, A. and 
	{Spurzem}, R.},
    title = "{Dynamical friction of massive objects in galactic centres}",
  journal = {\mnras},
archivePrefix = "arXiv",
   eprint = {1009.2455},
 primaryClass = "astro-ph.CO",
     year = 2011,
    month = feb,
   volume = 411,
    pages = {653-674},
      doi = {10.1111/j.1365-2966.2010.17711.x},
   adsurl = {http://adsabs.harvard.edu/abs/2011MNRAS.411..653J}
}

@ARTICLE{mer06,
   author = {{Merritt}, D.},
    title = "{Mass Deficits, Stalling Radii, and the Merger Histories of Elliptical Galaxies}",
  journal = {\apj},
   eprint = {astro-ph/0603439},
     year = 2006,
    month = sep,
   volume = 648,
    pages = {976-986},
      doi = {10.1086/506139},
   adsurl = {http://adsabs.harvard.edu/abs/2006ApJ...648..976M}
}

@ARTICLE{rantala+18,
   author = {{Rantala}, A. and {Johansson}, P.~H. and {Naab}, T. and {Thomas}, J. and 
	{Frigo}, M.},
    title = "{The Formation of Extremely Diffuse Galaxy Cores by Merging Supermassive Black Holes}",
  journal = {\apj},
archivePrefix = "arXiv",
   eprint = {1805.10295},
     year = 2018,
    month = sep,
   volume = 864,
      eid = {113},
    pages = {113},
      doi = {10.3847/1538-4357/aada47},
   adsurl = {http://adsabs.harvard.edu/abs/2018ApJ...864..113R}
}

@ARTICLE{bia19,
       author = {{Biava}, Nadia and {Colpi}, Monica and {Capelo}, Pedro R. and
         {Bonetti}, Matteo and {Volonteri}, Marta and {Tamfal}, Tomas and
         {Mayer}, Lucio and {Sesana}, Alberto},
        title = "{The lifetime of binary black holes in S{\'e}rsic galaxy models}",
      journal = {\mnras},
         year = "2019",
        month = "Jun",
        pages = {1542},
          doi = {10.1093/mnras/stz1614},
archivePrefix = {arXiv},
       eprint = {1903.05682},
 primaryClass = {astro-ph.GA},
       adsurl = {https://ui.adsabs.harvard.edu/abs/2019MNRAS.tmp.1542B}
}

@ARTICLE{spe17,
       author = {{Spengler}, Chelsea and {C{\^o}t{\'e}}, Patrick and {Roediger}, Joel and
         {Ferrarese}, Laura and {S{\'a}nchez-Janssen}, Rub{\'e}n and
         {Toloba}, Elisa and {Liu}, Yiqing and {Guhathakurta}, Puragra and
         {Cuillandre}, Jean-Charles and {Gwyn}, Stephen and {Zirm}, Andrew and
         {Mu{\~n}oz}, Roberto and {Puzia}, Thomas and {Lan{\c{c}}on}, Ariane and
         {Peng}, Eric W. and {Mei}, Simona and {Powalka}, Mathieu},
        title = "{Virgo Redux: The Masses and Stellar Content of Nuclei in Early-type Galaxies from Multiband Photometry and Spectroscopy}",
      journal = {\apj},
         year = "2017",
        month = "Nov",
       volume = {849},
       number = {1},
          eid = {55},
        pages = {55},
          doi = {10.3847/1538-4357/aa8a78},
archivePrefix = {arXiv},
       eprint = {1709.00406},
 primaryClass = {astro-ph.GA},
       adsurl = {https://ui.adsabs.harvard.edu/abs/2017ApJ...849...55S}
}

@ARTICLE{san19b,
       author = {{S{\'a}nchez-Janssen}, Rub{\'e}n and {C{\^o}t{\'e}}, Patrick and
         {Ferrarese}, Laura and {Peng}, Eric W. and {Roediger}, Joel and
         {Blakeslee}, John P. and {Emsellem}, Eric and {Puzia}, Thomas H. and
         {Spengler}, Chelsea and {Taylor}, James and
         {{\'A}lamo-Mart{\'\i}nez}, Karla A. and {Boselli}, Alessandro and
         {Cantiello}, Michele and {Cuillandre}, Jean-Charles and
         {Duc}, Pierre-Alain and {Durrell}, Patrick and {Gwyn}, Stephen and
         {MacArthur}, Lauren A. and {Lan{\c{c}}on}, Ariane and {Lim}, Sungsoon and
         {Liu}, Chengze and {Mei}, Simona and {Miller}, Bryan and
         {Mu{\~n}oz}, Roberto and {Mihos}, J. Christopher and {Paudel}, Sanjaya and
         {Powalka}, Mathieu and {Toloba}, Elisa},
        title = "{The Next Generation Virgo Cluster Survey. XXIII. Fundamentals of Nuclear Star Clusters over Seven Decades in Galaxy Mass}",
      journal = {\apj},
         year = "2019",
        month = "Jun",
       volume = {878},
       number = {1},
          eid = {18},
        pages = {18},
          doi = {10.3847/1538-4357/aaf4fd},
archivePrefix = {arXiv},
       eprint = {1812.01019},
 primaryClass = {astro-ph.GA},
       adsurl = {https://ui.adsabs.harvard.edu/abs/2019ApJ...878...18S}
}

@ARTICLE{Baile15,
   author = {{Li}, B. and {Holley-Bockelmann}, K. and {Khan}, F.~M.},
    title = "{Classification of Stellar Orbits in Axisymmetric Galaxies}",
  journal = {\apj},
archivePrefix = "arXiv",
   eprint = {1412.2134},
     year = 2015,
    month = sep,
   volume = 811,
      eid = {25},
    pages = {25},
      doi = {10.1088/0004-637X/811/1/25},
   adsurl = {http://adsabs.harvard.edu/abs/2015ApJ...811...25L}
}

@ARTICLE{geo14,
       author = {{Georgiev}, Iskren Y. and {B{\"o}ker}, Torsten},
        title = "{Nuclear star clusters in 228 spiral galaxies in the HST/WFPC2 archive: catalogue and comparison to other stellar systems}",
      journal = {\mnras},
         year = "2014",
        month = "Jul",
       volume = {441},
       number = {4},
        pages = {3570-3590},
          doi = {10.1093/mnras/stu797},
archivePrefix = {arXiv},
       eprint = {1404.5956},
 primaryClass = {astro-ph.GA},
       adsurl = {https://ui.adsabs.harvard.edu/abs/2014MNRAS.441.3570G}
}

@ARTICLE{bok04,
       author = {{B{\"o}ker}, Torsten and {Sarzi}, Marc and {McLaughlin}, Dean E. and
         {van der Marel}, Roeland P. and {Rix}, Hans-Walter and {Ho}, Luis C. and
         {Shields}, Joseph C.},
        title = "{A Hubble Space Telescope Census of Nuclear Star Clusters in Late-Type Spiral Galaxies. II. Cluster Sizes and Structural Parameter Correlations}",
      journal = {\aj},
         year = "2004",
        month = "Jan",
       volume = {127},
       number = {1},
        pages = {105-118},
          doi = {10.1086/380231},
archivePrefix = {arXiv},
       eprint = {astro-ph/0309761},
 primaryClass = {astro-ph},
       adsurl = {https://ui.adsabs.harvard.edu/abs/2004AJ....127..105B}
}

@ARTICLE{seth10,
       author = {{Seth}, Anil C. and {Cappellari}, Michele and {Neumayer}, Nadine and
         {Caldwell}, Nelson and {Bastian}, Nate and {Olsen}, Knut and
         {Blum}, Robert D. and {Debattista}, Victor P. and {McDermid}, Richard and
         {Puzia}, Thomas and {Stephens}, Andrew},
        title = "{The NGC 404 Nucleus: Star Cluster and Possible Intermediate-mass Black Hole}",
      journal = {\apj},
         year = "2010",
        month = "May",
       volume = {714},
       number = {1},
        pages = {713-731},
          doi = {10.1088/0004-637X/714/1/713},
archivePrefix = {arXiv},
       eprint = {1003.0680},
 primaryClass = {astro-ph.CO},
       adsurl = {https://ui.adsabs.harvard.edu/abs/2010ApJ...714..713S}
}

@ARTICLE{milo04,
       author = {{Milosavljevi{\'c}}, Milo{\v{s}}},
        title = "{On the Origin of Nuclear Star Clusters in Late-Type Spiral Galaxies}",
      journal = {\apjl},
         year = "2004",
        month = "Apr",
       volume = {605},
       number = {1},
        pages = {L13-L16},
          doi = {10.1086/420696},
archivePrefix = {arXiv},
       eprint = {astro-ph/0310574},
 primaryClass = {astro-ph},
       adsurl = {https://ui.adsabs.harvard.edu/abs/2004ApJ...605L..13M}
}

@ARTICLE{ant15,
       author = {{Antonini}, Fabio and {Barausse}, Enrico and {Silk}, Joseph},
        title = "{The Coevolution of Nuclear Star Clusters, Massive Black Holes, and Their Host Galaxies}",
      journal = {\apj},
         year = "2015",
        month = "Oct",
       volume = {812},
       number = {1},
          eid = {72},
        pages = {72},
          doi = {10.1088/0004-637X/812/1/72},
archivePrefix = {arXiv},
       eprint = {1506.02050},
 primaryClass = {astro-ph.GA},
       adsurl = {https://ui.adsabs.harvard.edu/abs/2015ApJ...812...72A}
}

@ARTICLE{ngu17,
       author = {{Nguyen}, Dieu D. and {Seth}, Anil C. and {den Brok}, Mark and
         {Neumayer}, Nadine and {Cappellari}, Michele and {Barth}, Aaron J. and
         {Caldwell}, Nelson and {Williams}, Benjamin F. and {Binder}, Breanna},
        title = "{Improved Dynamical Constraints on the Mass of the Central Black Hole in NGC 404}",
      journal = {\apj},
         year = "2017",
        month = "Feb",
       volume = {836},
       number = {2},
          eid = {237},
        pages = {237},
          doi = {10.3847/1538-4357/aa5cb4},
archivePrefix = {arXiv},
       eprint = {1610.09385},
 primaryClass = {astro-ph.GA},
       adsurl = {https://ui.adsabs.harvard.edu/abs/2017ApJ...836..237N}
}

@ARTICLE{neu20,
       author = {{Neumayer}, Nadine and {Seth}, Anil and {Boeker}, Torsten},
        title = "{Nuclear Star Clusters}",
      journal = {arXiv e-prints},
         year = "2020",
        month = "Jan",
          eid = {arXiv:2001.03626},
        pages = {arXiv:2001.03626},
archivePrefix = {arXiv},
       eprint = {2001.03626},
 primaryClass = {astro-ph.GA},
       adsurl = {https://ui.adsabs.harvard.edu/abs/2020arXiv200103626N}
}

@ARTICLE{tre75,
       author = {{Tremaine}, S.~D. and {Ostriker}, J.~P. and {Spitzer}, L., Jr.},
        title = "{The formation of the nuclei of galaxies. I. M31.}",
      journal = {\apj},
         year = "1975",
        month = "Mar",
       volume = {196},
        pages = {407-411},
          doi = {10.1086/153422},
       adsurl = {https://ui.adsabs.harvard.edu/abs/1975ApJ...196..407T}
}

@ARTICLE{ngu18,
       author = {{Nguyen}, Dieu D. and {Seth}, Anil C. and {Neumayer}, Nadine and
         {Kamann}, Sebastian and {Voggel}, Karina T. and {Cappellari}, Michele and
         {Picotti}, Arianna and {Nguyen}, Phuong M. and {B{\"o}ker}, Torsten and
         {Debattista}, Victor and {Caldwell}, Nelson and {McDermid}, Richard and
         {Bastian}, Nathan and {Ahn}, Christopher C. and {Pechetti}, Renuka},
        title = "{Nearby Early-type Galactic Nuclei at High Resolution: Dynamical Black Hole and Nuclear Star Cluster Mass Measurements}",
      journal = {\apj},
         year = "2018",
        month = "May",
       volume = {858},
       number = {2},
          eid = {118},
        pages = {118},
          doi = {10.3847/1538-4357/aabe28},
archivePrefix = {arXiv},
       eprint = {1711.04314},
 primaryClass = {astro-ph.GA},
       adsurl = {https://ui.adsabs.harvard.edu/abs/2018ApJ...858..118N}
}

@ARTICLE{ogi19,
       author = {{Ogiya}, Go and {Hahn}, Oliver and {Mingarelli}, Chiara M.~F. and
         {Volonteri}, Marta},
        title = "{Accelerated orbital decay of supermassive black hole binaries in merging nuclear star clusters}",
      journal = {arXiv e-prints},
         year = "2019",
        month = "Nov",
          eid = {arXiv:1911.11526},
        pages = {arXiv:1911.11526},
archivePrefix = {arXiv},
       eprint = {1911.11526},
 primaryClass = {astro-ph.GA},
       adsurl = {https://ui.adsabs.harvard.edu/abs/2019arXiv191111526O}
}

@BOOK{Merritt_2013,
       author = {{Merritt}, David},
        title = "{Dynamics and Evolution of Galactic Nuclei}",
         year = 2013,
    publisher = {Princeton: Princeton University Press},
       adsurl = {https://ui.adsabs.harvard.edu/abs/2013degn.book.....M}
}

@ARTICLE{Khan_Holley-Bockelmann2021,
       author = {{Khan}, Fazeel Mahmood and {Holley-Bockelmann}, Kelly},
        title = "{Extremely efficient mergers of intermediate-mass black hole binaries in nucleated dwarf galaxies}",
      journal = {\mnras},
         year = 2021,
        month = nov,
       volume = {508},
       number = {1},
        pages = {1174-1188},
          doi = {10.1093/mnras/stab2646},
archivePrefix = {arXiv},
       eprint = {2109.12129},
 primaryClass = {astro-ph.GA},
       adsurl = {https://ui.adsabs.harvard.edu/abs/2021MNRAS.508.1174K}
}

@ARTICLE{denbrok+2014,
       author = {{den Brok}, Mark and {Peletier}, Reynier F. and {Seth}, Anil and {Balcells}, Marc and {Dominguez}, Lilian and {Graham}, Alister W. and {Carter}, David and {Erwin}, Peter and {Ferguson}, Henry C. and {Goudfrooij}, Paul and {Guzm{\'a}n}, Rafael and {Hoyos}, Carlos and {Jogee}, Shardha and {Lucey}, John and {Phillipps}, Steven and {Puzia}, Thomas and {Valentijn}, Edwin and {Verdoes Kleijn}, Gijs and {Weinzirl}, Tim},
        title = "{The HST/ACS Coma Cluster Survey - X. Nuclear star clusters in low-mass early-type galaxies: scaling relations}",
      journal = {\mnras},
         year = 2014,
        month = dec,
       volume = {445},
       number = {3},
        pages = {2385-2403},
          doi = {10.1093/mnras/stu1906},
archivePrefix = {arXiv},
       eprint = {1409.4766},
 primaryClass = {astro-ph.GA},
       adsurl = {https://ui.adsabs.harvard.edu/abs/2014MNRAS.445.2385D}
}

@ARTICLE{Reines2022,
       author = {{Reines}, Amy E.},
        title = "{Hunting for massive black holes in dwarf galaxies}",
      journal = {Nature Astronomy},
         year = 2022,
        month = jan,
       volume = {6},
        pages = {26-34},
          doi = {10.1038/s41550-021-01556-0},
archivePrefix = {arXiv},
       eprint = {2201.10569},
 primaryClass = {astro-ph.GA},
       adsurl = {https://ui.adsabs.harvard.edu/abs/2022NatAs...6...26R}
}

@ARTICLE{mukh23,
       author = {{Mukherjee}, Diptajyoti and {Zhu}, Qirong and {Ogiya}, Go and {Rodriguez}, Carl L. and {Trac}, Hy},
        title = "{Evolution of massive black hole binaries in collisionally relaxed nuclear star clusters - Impact of mass segregation}",
      journal = {\mnras},
         year = 2023,
        month = feb,
       volume = {518},
       number = {4},
        pages = {4801-4817},
          doi = {10.1093/mnras/stac3390},
archivePrefix = {arXiv},
       eprint = {2205.12289},
 primaryClass = {astro-ph.GA},
       adsurl = {https://ui.adsabs.harvard.edu/abs/2023MNRAS.518.4801M}
}

@ARTICLE{ama23,
       author = {{Amaro-Seoane}, Pau and {Andrews}, Jeff and {Arca Sedda}, Manuel and {Askar}, Abbas and {Baghi}, Quentin and {Balasov}, Razvan and {Bartos}, Imre and {Bavera}, Simone S. and {Bellovary}, Jillian and {Berry}, Christopher P.~L. and {Berti}, Emanuele and {Bianchi}, Stefano and {Blecha}, Laura and {Blondin}, St{\'e}phane and {Bogdanovi{\'c}}, Tamara and {Boissier}, Samuel and {Bonetti}, Matteo and {Bonoli}, Silvia and {Bortolas}, Elisa and {Breivik}, Katelyn and {Capelo}, Pedro R. and {Caramete}, Laurentiu and {Cattorini}, Federico and {Charisi}, Maria and {Chaty}, Sylvain and {Chen}, Xian and {Chru{\'s}li{\'n}ska}, Martyna and {Chua}, Alvin J.~K. and {Church}, Ross and {Colpi}, Monica and {D'Orazio}, Daniel and {Danielski}, Camilla and {Davies}, Melvyn B. and {Dayal}, Pratika and {De Rosa}, Alessandra and {Derdzinski}, Andrea and {Destounis}, Kyriakos and {Dotti}, Massimo and {Du{\r{A}}{\textsterling}an}, Ioana and {Dvorkin}, Irina and {Fabj}, Gaia and {Foglizzo}, Thierry and {Ford}, Saavik and {Fouvry}, Jean-Baptiste and {Franchini}, Alessia and {Fragos}, Tassos and {Fryer}, Chris and {Gaspari}, Massimo and {Gerosa}, Davide and {Graziani}, Luca and {Groot}, Paul and {Habouzit}, Melanie and {Haggard}, Daryl and {Haiman}, Zoltan and {Han}, Wen-Biao and {Istrate}, Alina and {Johansson}, Peter H. and {Khan}, Fazeel Mahmood and {Kimpson}, Tomas and {Kokkotas}, Kostas and {Kong}, Albert and {Korol}, Valeriya and {Kremer}, Kyle and {Kupfer}, Thomas and {Lamberts}, Astrid and {Larson}, Shane and {Lau}, Mike and {Liu}, Dongliang and {Lloyd-Ronning}, Nicole and {Lodato}, Giuseppe and {Lupi}, Alessandro and {Ma}, Chung-Pei and {Maccarone}, Tomas and {Mandel}, Ilya and {Mangiagli}, Alberto and {Mapelli}, Michela and {Mathis}, St{\'e}phane and {Mayer}, Lucio and {McGee}, Sean and {McKernan}, Barry and {Miller}, M. Coleman and {Mota}, David F. and {Mumpower}, Matthew and {Nasim}, Syeda S. and {Nelemans}, Gijs and {Noble}, Scott and {Pacucci}, Fabio and {Panessa}, Francesca and {Paschalidis}, Vasileios and {Pfister}, Hugo and {Porquet}, Delphine and {Quenby}, John and {Ricarte}, Angelo and {R{\"o}pke}, Friedrich K. and {Regan}, John and {Rosswog}, Stephan and {Ruiter}, Ashley and {Ruiz}, Milton and {Runnoe}, Jessie and {Schneider}, Raffaella and {Schnittman}, Jeremy and {Secunda}, Amy and {Sesana}, Alberto and {Seto}, Naoki and {Shao}, Lijing and {Shapiro}, Stuart and {Sopuerta}, Carlos and {Stone}, Nicholas C. and {Suvorov}, Arthur and {Tamanini}, Nicola and {Tamfal}, Tomas and {Tauris}, Thomas and {Temmink}, Karel and {Tomsick}, John and {Toonen}, Silvia and {Torres-Orjuela}, Alejandro and {Toscani}, Martina and {Tsokaros}, Antonios and {Unal}, Caner and {V{\'a}zquez-Aceves}, Ver{\'o}nica and {Valiante}, Rosa and {van Putten}, Maurice and {van Roestel}, Jan and {Vignali}, Christian and {Volonteri}, Marta and {Wu}, Kinwah and {Younsi}, Ziri and {Yu}, Shenghua and {Zane}, Silvia and {Zwick}, Lorenz and {Antonini}, Fabio and {Baibhav}, Vishal and {Barausse}, Enrico and {Bonilla Rivera}, Alexander and {Branchesi}, Marica and {Branduardi-Raymont}, Graziella and {Burdge}, Kevin and {Chakraborty}, Srija and {Cuadra}, Jorge and {Dage}, Kristen and {Davis}, Benjamin and {de Mink}, Selma E. and {Decarli}, Roberto and {Doneva}, Daniela and {Escoffier}, Stephanie and {Gandhi}, Poshak and {Haardt}, Francesco and {Lousto}, Carlos O. and {Nissanke}, Samaya and {Nordhaus}, Jason and {O'Shaughnessy}, Richard and {Portegies Zwart}, Simon and {Pound}, Adam and {Schussler}, Fabian and {Sergijenko}, Olga and {Spallicci}, Alessandro and {Vernieri}, Daniele and {Vigna-G{\'o}mez}, Alejandro},
        title = "{Astrophysics with the Laser Interferometer Space Antenna}",
      journal = {Living Reviews in Relativity},
         year = 2023,
        month = dec,
       volume = {26},
       number = {1},
          eid = {2},
        pages = {2},
          doi = {10.1007/s41114-022-00041-y},
archivePrefix = {arXiv},
       eprint = {2203.06016},
 primaryClass = {gr-qc},
       adsurl = {https://ui.adsabs.harvard.edu/abs/2023LRR....26....2A}
}

@article{JUWELS,
author = {{J\"{u}lich Supercomputing Centre}},
title = {{JUWELS Cluster and Booster: Exascale Pathfinder with Modular Supercomputing Architecture at Juelich Supercomputing Centre}},
journal = {Journal of large-scale research facilities},
number = {A138},
volume = {7},
doi = {10.17815/jlsrf-7-183},
year = {2021}
}

@ARTICLE{bald17,
       author = {{Baldassare}, Vivienne F. and {Reines}, Amy E. and {Gallo}, Elena and {Greene}, Jenny E.},
        title = "{X-ray and Ultraviolet Properties of AGNs in Nearby Dwarf Galaxies}",
      journal = {\apj},
         year = 2017,
        month = feb,
       volume = {836},
       number = {1},
          eid = {20},
        pages = {20},
          doi = {10.3847/1538-4357/836/1/20},
archivePrefix = {arXiv},
       eprint = {1609.07148},
 primaryClass = {astro-ph.HE},
       adsurl = {https://ui.adsabs.harvard.edu/abs/2017ApJ...836...20B}
}

@ARTICLE{bald20,
       author = {{Baldassare}, Vivienne F. and {Geha}, Marla and {Greene}, Jenny},
        title = "{A Search for Optical AGN Variability in 35,000 Low-mass Galaxies with the Palomar Transient Factory}",
      journal = {\apj},
         year = 2020,
        month = jun,
       volume = {896},
       number = {1},
          eid = {10},
        pages = {10},
          doi = {10.3847/1538-4357/ab8936},
archivePrefix = {arXiv},
       eprint = {1910.06342},
 primaryClass = {astro-ph.HE},
       adsurl = {https://ui.adsabs.harvard.edu/abs/2020ApJ...896...10B}
}

@ARTICLE{fre20,
       author = {{French}, K. Decker and {Wevers}, Thomas and {Law-Smith}, Jamie and {Graur}, Or and {Zabludoff}, Ann I.},
        title = "{The Host Galaxies of Tidal Disruption Events}",
      journal = {\ssr},
         year = 2020,
        month = mar,
       volume = {216},
       number = {3},
          eid = {32},
        pages = {32},
          doi = {10.1007/s11214-020-00657-y},
archivePrefix = {arXiv},
       eprint = {2003.02863},
 primaryClass = {astro-ph.HE},
       adsurl = {https://ui.adsabs.harvard.edu/abs/2020SSRv..216...32F}
}

@ARTICLE{mer09,
       author = {{Merritt}, David},
        title = "{Evolution of Nuclear Star Clusters}",
      journal = {\apj},
         year = 2009,
        month = apr,
       volume = {694},
       number = {2},
        pages = {959-970},
          doi = {10.1088/0004-637X/694/2/959},
archivePrefix = {arXiv},
       eprint = {0802.3186},
 primaryClass = {astro-ph},
       adsurl = {https://ui.adsabs.harvard.edu/abs/2009ApJ...694..959M}
}

@ARTICLE{sturm24,
       author = {{Sturm}, Megan R. and {Reines}, Amy E.},
        title = "{A Breakdown of the Black Hole{\textendash}Bulge Mass Relation in Local Active Galaxies}",
      journal = {\apj},
         year = 2024,
        month = aug,
       volume = {971},
       number = {2},
          eid = {173},
        pages = {173},
          doi = {10.3847/1538-4357/ad55e6},
archivePrefix = {arXiv},
       eprint = {2406.06675},
 primaryClass = {astro-ph.GA},
       adsurl = {https://ui.adsabs.harvard.edu/abs/2024ApJ...971..173S}
}

@ARTICLE{fah22,
       author = {{Fahrion}, Katja and {Leaman}, Ryan and {Lyubenova}, Mariya and {van de Ven}, Glenn},
        title = "{Disentangling the formation mechanisms of nuclear star clusters}",
      journal = {\aap},
         year = 2022,
        month = feb,
       volume = {658},
          eid = {A172},
        pages = {A172},
          doi = {10.1051/0004-6361/202039778},
archivePrefix = {arXiv},
       eprint = {2112.05610},
 primaryClass = {astro-ph.GA},
       adsurl = {https://ui.adsabs.harvard.edu/abs/2022A&A...658A.172F}
}

@ARTICLE{cap93,
       author = {{Capuzzo-Dolcetta}, Roberto},
        title = "{The Evolution of the Globular Cluster System in a Triaxial Galaxy: Can a Galactic Nucleus Form by Globular Cluster Capture?}",
      journal = {\apj},
         year = 1993,
        month = oct,
       volume = {415},
        pages = {616},
          doi = {10.1086/173189},
archivePrefix = {arXiv},
       eprint = {astro-ph/9301006},
 primaryClass = {astro-ph},
       adsurl = {https://ui.adsabs.harvard.edu/abs/1993ApJ...415..616C}
}

@ARTICLE{colpi24,
       author = {{Colpi}, Monica and {Danzmann}, Karsten and {Hewitson}, Martin and {Holley-Bockelmann}, Kelly and {Jetzer}, Philippe and {Nelemans}, Gijs and {Petiteau}, Antoine and {Shoemaker}, David and {Sopuerta}, Carlos and {Stebbins}, Robin and {Tanvir}, Nial and {Ward}, Henry and {Weber}, William Joseph and {Thorpe}, Ira and {Daurskikh}, Anna and {Deep}, Atul and {Fern{\'a}ndez N{\'u}{\~n}ez}, Ignacio and {Garc{\'\i}a Marirrodriga}, C{\'e}sar and {Gehler}, Martin and {Halain}, Jean-Philippe and {Jennrich}, Oliver and {Lammers}, Uwe and {Larra{\~n}aga}, Jonan and {Lieser}, Maike and {L{\"u}tzgendorf}, Nora and {Martens}, Waldemar and {Mondin}, Linda and {Piris Ni{\~n}o}, Ana and {Amaro-Seoane}, Pau and {Arca Sedda}, Manuel and {Auclair}, Pierre and {Babak}, Stanislav and {Baghi}, Quentin and {Baibhav}, Vishal and {Baker}, Tessa and {Bayle}, Jean-Baptiste and {Berry}, Christopher and {Berti}, Emanuele and {Boileau}, Guillaume and {Bonetti}, Matteo and {Brito}, Richard and {Buscicchio}, Riccardo and {Calcagni}, Gianluca and {Capelo}, Pedro R. and {Caprini}, Chiara and {Caputo}, Andrea and {Castelli}, Eleonora and {Chen}, Hsin-Yu and {Chen}, Xian and {Chua}, Alvin and {Davies}, Gareth and {Derdzinski}, Andrea and {Domcke}, Valerie Fiona and {Doneva}, Daniela and {Dvorkin}, Irna and {Mar{\'\i}a Ezquiaga}, Jose and {Gair}, Jonathan and {Haiman}, Zoltan and {Harry}, Ian and {Hartwig}, Olaf and {Hees}, Aurelien and {Heffernan}, Anna and {Husa}, Sascha and {Izquierdo-Villalba}, David and {Karnesis}, Nikolaos and {Klein}, Antoine and {Korol}, Valeriya and {Korsakova}, Natalia and {Kupfer}, Thomas and {Laghi}, Danny and {Lamberts}, Astrid and {Larson}, Shane and {Le Jeune}, Maude and {Lewicki}, Marek and {Littenberg}, Tyson and {Madge}, Eric and {Mangiagli}, Alberto and {Marsat}, Sylvain and {Vilchez}, Ivan Martin and {Maselli}, Andrea and {Mathews}, Josh and {van de Meent}, Maarten and {Muratore}, Martina and {Nardini}, Germano and {Pani}, Paolo and {Peloso}, Marco and {Pieroni}, Mauro and {Pound}, Adam and {Quelquejay-Leclere}, Hippolyte and {Ricciardone}, Angelo and {Rossi}, Elena Maria and {Sartirana}, Andrea and {Savalle}, Etienne and {Sberna}, Laura and {Sesana}, Alberto and {Shoemaker}, Deirdre and {Slutsky}, Jacob and {Sotiriou}, Thomas and {Speri}, Lorenzo and {Staab}, Martin and {Steer}, Dani{\`e}le and {Tamanini}, Nicola and {Tasinato}, Gianmassimo and {Torrado}, Jesus and {Torres-Orjuela}, Alejandro and {Toubiana}, Alexandre and {Vallisneri}, Michele and {Vecchio}, Alberto and {Volonteri}, Marta and {Yagi}, Kent and {Zwick}, Lorenz},
        title = "{LISA Definition Study Report}",
      journal = {arXiv e-prints},
         year = 2024,
        month = feb,
          eid = {arXiv:2402.07571},
        pages = {arXiv:2402.07571},
          doi = {10.48550/arXiv.2402.07571},
archivePrefix = {arXiv},
       eprint = {2402.07571},
 primaryClass = {astro-ph.CO},
       adsurl = {https://ui.adsabs.harvard.edu/abs/2024arXiv240207571C}
}

@ARTICLE{pan19,
       author = {{Panamarev}, Taras and {Just}, Andreas and {Spurzem}, Rainer and {Berczik}, Peter and {Wang}, Long and {Arca Sedda}, Manuel},
        title = "{Direct N-body simulation of the Galactic centre}",
      journal = {\mnras},
         year = 2019,
        month = apr,
       volume = {484},
       number = {3},
        pages = {3279-3290},
          doi = {10.1093/mnras/stz208},
archivePrefix = {arXiv},
       eprint = {1805.02153},
 primaryClass = {astro-ph.GA},
       adsurl = {https://ui.adsabs.harvard.edu/abs/2019MNRAS.484.3279P}
}

@ARTICLE{sobo22,
       author = {{Sobolenko}, M. and {Kompaniiets}, O. and {Berczik}, P. and {Marchenko}, V. and {Vasylenko}, A. and {Fedorova}, E. and {Shukirgaliyev}, B.},
        title = "{NGC 6240 supermassive black hole binary dynamical evolution based on Chandra data}",
      journal = {\mnras},
         year = 2022,
        month = dec,
       volume = {517},
       number = {2},
        pages = {1791-1802},
          doi = {10.1093/mnras/stac2472},
archivePrefix = {arXiv},
       eprint = {2209.01264},
 primaryClass = {astro-ph.GA},
       adsurl = {https://ui.adsabs.harvard.edu/abs/2022MNRAS.517.1791S}
}

@ARTICLE{ber22,
       author = {{Berczik}, Peter and {Arca Sedda}, Manuel and {Sobolenko}, Margaryta and {Ishchenko}, Marina and {Sobodar}, Olexander and {Spurzem}, Rainer},
        title = "{Merging of unequal mass binary black holes in non-axisymmetric galactic nuclei}",
      journal = {\aap},
         year = 2022,
        month = sep,
       volume = {665},
          eid = {A86},
        pages = {A86},
          doi = {10.1051/0004-6361/202244354},
archivePrefix = {arXiv},
       eprint = {2008.04342},
 primaryClass = {astro-ph.GA},
       adsurl = {https://ui.adsabs.harvard.edu/abs/2022A&A...665A..86B}
}

@ARTICLE{mukherjee24,
       author = {{Mukherjee}, Diptajyoti and {Zhou}, Yihao and {Chen}, Nianyi and {Di Carlo}, Ugo Niccol{\'o} and {Di Matteo}, Tiziana},
        title = "{MAGICS III. Seeds sink swiftly: nuclear star clusters dramatically accelerate seed black hole mergers}",
      journal = {arXiv e-prints},
         year = 2024,
        month = sep,
          eid = {arXiv:2409.19095},
        pages = {arXiv:2409.19095},
          doi = {10.48550/arXiv.2409.19095},
archivePrefix = {arXiv},
       eprint = {2409.19095},
 primaryClass = {astro-ph.GA},
       adsurl = {https://ui.adsabs.harvard.edu/abs/2024arXiv240919095M}
}

@ARTICLE{khan+24,
       author = {{Khan}, Fazeel Mahmood and {Javed}, Fiza and {Holley-Bockelmann}, Kelly and {Mayer}, Lucio and {Berczik}, Peter and {Macci{\`o}}, Andrea V.},
        title = "{The Potential for Long-lived Intermediate-mass Black Hole Binaries in the Lowest Density Dwarf Galaxies}",
      journal = {\apj},
         year = 2024,
        month = nov,
       volume = {976},
       number = {1},
          eid = {22},
        pages = {22},
          doi = {10.3847/1538-4357/ad8082},
archivePrefix = {arXiv},
       eprint = {2408.14541},
 primaryClass = {astro-ph.GA},
       adsurl = {https://ui.adsabs.harvard.edu/abs/2024ApJ...976...22K}
}

@ARTICLE{zhou25,
       author = {{Zhou}, Yihao and {Mukherjee}, Diptajyoti and {Chen}, Nianyi and {Di Matteo}, Tiziana and {Johansson}, Peter H. and {Rantala}, Antti and {Partmann}, Christian and {Di Carlo}, Ugo Niccol{\`o} and {Bird}, Simeon and {Ni}, Yueying},
        title = "{MAGICS. II. Seed Black Holes Stripped of Their Surrounding Stars Do Not Sink}",
      journal = {\apj},
         year = 2025,
        month = feb,
       volume = {980},
       number = {1},
          eid = {79},
        pages = {79},
          doi = {10.3847/1538-4357/ada283},
archivePrefix = {arXiv},
       eprint = {2409.19914},
 primaryClass = {astro-ph.GA},
       adsurl = {https://ui.adsabs.harvard.edu/abs/2025ApJ...980...79Z}
}

@ARTICLE{ArcaSeddaetal2024,
       author = {{Arca Sedda}, Manuel and {Kamlah}, Albrecht W.~H. and {Spurzem}, Rainer and {Rizzuto}, Francesco Paolo and {Giersz}, Mirek and {Naab}, Thorsten and {Berczik}, Peter},
        title = "{The DRAGON-II simulations - III. Compact binary mergers in clusters with up to 1 million stars: mass, spin, eccentricity, merger rate, and pair instability supernovae rate}",
      journal = {\mnras},
         year = 2024,
        month = mar,
       volume = {528},
       number = {3},
        pages = {5140-5159},
          doi = {10.1093/mnras/stad3951},
archivePrefix = {arXiv},
       eprint = {2307.04807},
 primaryClass = {astro-ph.HE},
       adsurl = {https://ui.adsabs.harvard.edu/abs/2024MNRAS.528.5140A}
}

@ARTICLE{RantalaNaab2025,
       author = {{Rantala}, Antti and {Naab}, Thorsten},
        title = "{A rapid channel for the collisional formation and gravitational wave-driven mergers of supermassive black hole seeds at high redshift}",
      journal = {\mnras},
         year = 2025,
        month = sep,
       volume = {542},
       number = {1},
        pages = {L78-L84},
          doi = {10.1093/mnrasl/slaf064},
       adsurl = {https://ui.adsabs.harvard.edu/abs/2025MNRAS.542L..78R}
}

@ARTICLE{Rizzutoetal2023,
       author = {{Rizzuto}, Francesco Paolo and {Naab}, Thorsten and {Rantala}, Antti and {Johansson}, Peter H. and {Ostriker}, Jeremiah P. and {Stone}, Nicholas C. and {Liao}, Shihong and {Irodotou}, Dimitrios},
        title = "{The growth of intermediate mass black holes through tidal captures and tidal disruption events}",
      journal = {\mnras},
         year = 2023,
        month = may,
       volume = {521},
       number = {2},
        pages = {2930-2948},
          doi = {10.1093/mnras/stad734},
archivePrefix = {arXiv},
       eprint = {2211.13320},
 primaryClass = {astro-ph.GA},
       adsurl = {https://ui.adsabs.harvard.edu/abs/2023MNRAS.521.2930R}
}

@ARTICLE{Rizzutoetal2022,
       author = {{Rizzuto}, Francesco Paolo and {Naab}, Thorsten and {Spurzem}, Rainer and {Arca-Sedda}, Manuel and {Giersz}, Mirek and {Ostriker}, Jeremiah Paul and {Banerjee}, Sambaran},
        title = "{Black hole mergers in compact star clusters and massive black hole formation beyond the mass gap}",
      journal = {\mnras},
         year = 2022,
        month = may,
       volume = {512},
       number = {1},
        pages = {884-898},
          doi = {10.1093/mnras/stac231},
archivePrefix = {arXiv},
       eprint = {2108.11457},
 primaryClass = {astro-ph.GA},
       adsurl = {https://ui.adsabs.harvard.edu/abs/2022MNRAS.512..884R}
}

@ARTICLE{Vergaraetal2025,
       author = {{Vergara}, Marcelo C. and {Askar}, Abbas and {Kamlah}, Albrecht W.~H. and {Spurzem}, Rainer and {Flammini Dotti}, Francesco and {Schleicher}, Dominik R.~G. and {Arca Sedda}, Manuel and {Hypki}, Arkadiusz and {Giersz}, Mirek and {Hurley}, Jarrod and {Berczik}, Peter and {Escala}, Andres and {Hoyer}, Nils and {Neumayer}, Nadine and {Pang}, Xiaoying and {Tanikawa}, Ataru and {Cen}, Renyue and {Naab}, Thorsten},
        title = "{Rapid formation of a very massive star >50000 $M_\odot$ and subsequently an IMBH from runaway collisions. Direct N-body and Monte Carlo simulations of dense star clusters}",
      journal = {arXiv e-prints},
         year = 2025,
        month = may,
          eid = {arXiv:2505.07491},
        pages = {arXiv:2505.07491},
          doi = {10.48550/arXiv.2505.07491},
archivePrefix = {arXiv},
       eprint = {2505.07491},
 primaryClass = {astro-ph.GA},
       adsurl = {https://ui.adsabs.harvard.edu/abs/2025arXiv250507491V}
}

@ARTICLE{Vergaraetal2024,
       author = {{Vergara}, M.~C. and {Schleicher}, D.~R.~G. and {Escala}, A. and {Reinoso}, B. and {Flammini Dotti}, F. and {Kamlah}, A.~W.~H. and {Liempi}, M. and {Hoyer}, N. and {Neumayer}, N. and {Spurzem}, R.},
        title = "{Efficiency of black hole formation via collisions in stellar systems: Data analysis from simulations and observations}",
      journal = {\aap},
         year = 2024,
        month = sep,
       volume = {689},
          eid = {A34},
        pages = {A34},
          doi = {10.1051/0004-6361/202449967},
archivePrefix = {arXiv},
       eprint = {2405.12008},
 primaryClass = {astro-ph.GA},
       adsurl = {https://ui.adsabs.harvard.edu/abs/2024A&A...689A..34V}
}

@ARTICLE{McCaffrey2025,
       author = {{McCaffrey}, Joe and {Regan}, John and {Smith}, Britton and {Wise}, John and {O'Shea}, Brian and {Norman}, Michael},
        title = "{A Heavy Seed Black Hole Mass Function at High Redshift {\textendash} Prospects for LISA}",
      journal = {The Open Journal of Astrophysics},
         year = 2025,
        month = jan,
       volume = {8},
          eid = {11},
        pages = {11},
          doi = {10.33232/001c.129138},
archivePrefix = {arXiv},
       eprint = {2409.16413},
 primaryClass = {astro-ph.GA},
       adsurl = {https://ui.adsabs.harvard.edu/abs/2025OJAp....8E..11M}
}

@ARTICLE{Sathyaprakashetal2012,
       author = {{Sathyaprakash}, B. and {Abernathy}, M. and {Acernese}, F. and {Ajith}, P. and {Allen}, B. and {Amaro-Seoane}, P. and {Andersson}, N. and {Aoudia}, S. and {Arun}, K. and {Astone}, P. and {Krishnan}, B. and {Barack}, L. and {Barone}, F. and {Barr}, B. and {Barsuglia}, M. and {Bassan}, M. and {Bassiri}, R. and {Beker}, M. and {Beveridge}, N. and {Bizouard}, M. and {Bond}, C. and {Bose}, S. and {Bosi}, L. and {Braccini}, S. and {Bradaschia}, C. and {Britzger}, M. and {Brueckner}, F. and {Bulik}, T. and {Bulten}, H.~J. and {Burmeister}, O. and {Calloni}, E. and {Campsie}, P. and {Carbone}, L. and {Cella}, G. and {Chalkley}, E. and {Chassande-Mottin}, E. and {Chelkowski}, S. and {Chincarini}, A. and {Di Cintio}, A. and {Clark}, J. and {Coccia}, E. and {Colacino}, C.~N. and {Colas}, J. and {Colla}, A. and {Corsi}, A. and {Cumming}, A. and {Cunningham}, L. and {Cuoco}, E. and {Danilishin}, S. and {Danzmann}, K. and {Daw}, E. and {De Salvo}, R. and {Del Pozzo}, W. and {Dent}, T. and {De Rosa}, R. and {Di Fiore}, L. and {Emilio}, M. Di Paolo and {Di Virgilio}, A. and {Dietz}, A. and {Doets}, M. and {Dueck}, J. and {Edwards}, M. and {Fafone}, V. and {Fairhurst}, S. and {Falferi}, P. and {Favata}, M. and {Ferrari}, V. and {Ferrini}, F. and {Fidecaro}, F. and {Flaminio}, R. and {Franc}, J. and {Frasconi}, F. and {Freise}, A. and {Friedrich}, D. and {Fulda}, P. and {Gair}, J. and {Galimberti}, M. and {Gemme}, G. and {Genin}, E. and {Gennai}, A. and {Giazotto}, A. and {Glampedakis}, K. and {Gossan}, S. and {Gouaty}, R. and {Graef}, C. and {Graham}, W. and {Granata}, M. and {Grote}, H. and {Guidi}, G. and {Hallam}, J. and {Hammond}, G. and {Hannam}, M. and {Harms}, J. and {Haughian}, K. and {Hawke}, I. and {Heinert}, D. and {Hendry}, M. and {Heng}, I. and {Hennes}, E. and {Hild}, S. and {Hough}, J. and {Huet}, D. and {Husa}, S. and {Huttner}, S. and {Iyer}, B. and {Jones}, D.~I. and {Jones}, G. and {Kamaretsos}, I. and {Kant Mishra}, C. and {Kawazoe}, F. and {Khalili}, F. and {Kley}, B. and {Kokeyama}, K. and {Kokkotas}, K. and {Kroker}, S. and {Kumar}, R. and {Kuroda}, K. and {Lagrange}, B. and {Lastzka}, N. and {Li}, T.~G.~F. and {Lorenzini}, M. and {Losurdo}, G. and {L{\"u}ck}, H. and {Majorana}, E. and {Malvezzi}, V. and {Mandel}, I. and {Mandic}, V. and {Marka}, S. and {Marin}, F. and {Marion}, F. and {Marque}, J. and {Martin}, I. and {McLeod}, D. and {Mckechan}, D. and {Mehmet}, M. and {Michel}, C. and {Minenkov}, Y. and {Morgado}, N. and {Morgia}, A. and {Mosca}, S. and {Moscatelli}, L. and {Mours}, B. and {M{\"u}ller-Ebhardt}, H. and {Murray}, P. and {Naticchioni}, L. and {Nawrodt}, R. and {Nelson}, J. and {O' Shaughnessy}, R. and {Ott}, C.~D. and {Palomba}, C. and {Paoli}, A. and {Parguez}, G. and {Pasqualetti}, A. and {Passaquieti}, R. and {Passuello}, D. and {Perciballi}, M. and {Piergiovanni}, F. and {Pinard}, L. and {Pitkin}, M. and {Plastino}, W. and {Plissi}, M. and {Poggiani}, R. and {Popolizio}, P. and {Porter}, E. and {Prato}, M. and {Prodi}, G. and {Punturo}, M. and {Puppo}, P. and {Rabeling}, D. and {Racz}, I. and {Rapagnani}, P. and {Re}, V. and {Read}, J. and {Regimbau}, T. and {Rehbein}, H. and {Reid}, S. and {Ricci}, F. and {Richard}, F. and {Robinson}, C. and {Rocchi}, A. and {Romano}, R. and {Rowan}, S. and {R{\"u}diger}, A. and {Samblowski}, A. and {Santamar{\'\i}a}, L. and {Sassolas}, B. and {Schilling}, R. and {Schmidt}, P. and {Schnabel}, R. and {Schutz}, B. and {Schwarz}, C. and {Scott}, J. and {Seidel}, P. and {Sintes}, A.~M. and {Somiya}, K. and {Sopuerta}, C.~F. and {Sorazu}, B. and {Speirits}, F. and {Storchi}, L. and {Strain}, K.},
        title = "{Scientific objectives of Einstein Telescope}",
      journal = {Classical and Quantum Gravity},
         year = 2012,
        month = jun,
       volume = {29},
       number = {12},
          eid = {124013},
        pages = {124013},
          doi = {10.1088/0264-9381/29/12/124013},
archivePrefix = {arXiv},
       eprint = {1206.0331},
 primaryClass = {gr-qc},
       adsurl = {https://ui.adsabs.harvard.edu/abs/2012CQGra..29l4013S}
}

@ARTICLE{DiGiovanni2025,
       author = {{Di Giovanni}, Matteo},
        title = "{Einstein Telescope and Cosmic Explorer}",
      journal = {arXiv e-prints},
         year = 2025,
        month = may,
          eid = {arXiv:2505.11033},
        pages = {arXiv:2505.11033},
          doi = {10.48550/arXiv.2505.11033},
archivePrefix = {arXiv},
       eprint = {2505.11033},
 primaryClass = {gr-qc},
       adsurl = {https://ui.adsabs.harvard.edu/abs/2025arXiv250511033D}
}

@ARTICLE{Abacetal2025,
       author = {{Abac}, Adrian and {Abramo}, Raul and {Albanesi}, Simone and {Albertini}, Angelica and {Agapito}, Alessandro and {Agathos}, Michalis and {Albertus}, Conrado and {Andersson}, Nils and {Andrade}, Tom{\'a}s and {Andreoni}, Igor and {Angeloni}, Federico and {Antonelli}, Marco and {Antoniadis}, John and {Antonini}, Fabio and {Arca Sedda}, Manuel and {Artale}, M. Celeste and {Ascenzi}, Stefano and {Auclair}, Pierre and {Bachetti}, Matteo and {Badger}, Charles and {Banerjee}, Biswajit and {Barba-Gonz{\'a}lez}, David and {Barta}, D{\'a}niel and {Bartolo}, Nicola and {Bauswein}, Andreas and {Begnoni}, Andrea and {Beirnaert}, Freija and {Bejger}, Micha{\l} and {Belgacem}, Enis and {Bellomo}, Nicola and {Bernard}, Laura and {Grazia Bernardini}, Maria and {Bernuzzi}, Sebastiano and {Berry}, Christopher P.~L. and {Berti}, Emanuele and {Bertone}, Gianfranco and {Bettoni}, Dario and {Bezares}, Miguel and {Bhagwat}, Swetha and {Bisero}, Sofia and {Bizouard}, Marie Anne and {Blanco-Pillado}, Jose J. and {Blasi}, Simone and {Bonino}, Alice and {Borghese}, Alice and {Borghi}, Nicola and {Borhanian}, Ssohrab and {Bortolas}, Elisa and {Botticella}, Maria Teresa and {Branchesi}, Marica and {Breschi}, Matteo and {Brito}, Richard and {Brocato}, Enzo and {Broekgaarden}, Floor S. and {Bulik}, Tomasz and {Buonanno}, Alessandra and {Burgio}, Fiorella and {Burrows}, Adam and {Calcagni}, Gianluca and {Canevarolo}, Sofia and {Cappellaro}, Enrico and {Capurri}, Giulia and {Carbone}, Carmelita and {Casadio}, Roberto and {Cayuso}, Ramiro and {Cerd{\'a}-Dur{\'a}n}, Pablo and {Char}, Prasanta and {Chaty}, Sylvain and {Chiarusi}, Tommaso and {Chruslinska}, Martyna and {Cireddu}, Francesco and {Cole}, Philippa and {Colombo}, Alberto and {Colpi}, Monica and {Comp{\`e}re}, Geoffrey and {Contaldi}, Carlo and {Corman}, Maxence and {Crescimbeni}, Francesco and {Cristallo}, Sergio and {Cuoco}, Elena and {Cusin}, Giulia and {Dal Canton}, Tito and {D{\'a}lya}, Gergely and {D'Avanzo}, Paolo and {Davari}, Nazanin and {De Luca}, Valerio and {De Renzis}, Viola and {Della Valle}, Massimo and {Del Pozzo}, Walter and {De Santi}, Federico and {Ludovico De Santis}, Alessio and {Dietrich}, Tim and {Dimastrogiovanni}, Ema and {Domenech}, Guillem and {Doneva}, Daniela and {Drago}, Marco and {Dupletsa}, Ulyana and {Duval}, Hannah and {Dvorkin}, Irina and {Elias-Rosa}, Nancy and {Fairhurst}, Stephen and {Fantina}, Anthea F. and {Fasiello}, Matteo and {Fays}, Maxime and {Fender}, Rob and {Fischer}, Tobias and {Foucart}, Fran{\c{c}}ois and {Fragos}, Tassos and {Foffa}, Stefano and {Franciolini}, Gabriele and {Gair}, Jonathan and {Gamba}, Rossella and {Garcia-Bellido}, Juan and {Garc{\'\i}a-Quir{\'o}s}, Cecilio and {{\'A}rp{\'a}d Gergely}, L{\'a}szl{\'o} and {Ghirlanda}, Giancarlo and {Ghosh}, Archisman and {Giacomazzo}, Bruno and {Gittins}, Fabian and {Giudice}, Ines Francesca and {Goncharov}, Boris and {Gonzalez}, Alejandra and {Gori{\'e}ly}, St{\'e}phane and {Graziani}, Luca and {Greco}, Giuseppe and {Gualtieri}, Leonardo and {Guidi}, Gianluca Maria and {Gupta}, Ish and {Haney}, Maria and {Hannam}, Mark and {Harms}, Jan and {Harutyunyan}, Arus and {Haskell}, Brynmor and {Haungs}, Andreas and {Hazra}, Nandini and {Hemming}, Gary and {Heng}, Ik Siong and {Hinderer}, Tanja and {van der Horst}, Alexander and {Hu}, Qian and {Husa}, Sascha and {Iacovelli}, Francesco and {Illuminati}, Giulia and {Inguglia}, Gianluca and {Izquierdo Villalba}, David and {Janquart}, Justin and {Janssens}, Kamiel and {Jenkins}, Alexander C. and {Jones}, Ian and {Kacskovics}, Bal{\'a}zs and {Klessen}, Ralf S. and {Kokkotas}, Kostas and {Kuan}, Hao-Jui and {Kumar}, Sumit and {Kuroyanagi}, Sachiko and {Laghi}, Danny and {Lamberts}, Astrid and {Lambiase}, Gaetano and {Larrouturou}, Fran{\c{c}}ois and {Leaci}, Paola and {Lenzi}, Michele and {Levan}, Andrew and {Li}, T.~G.~F. and {Li}, Yufeng and {Liang}, Dicong and {Limongi}, Marco and {Liu}, Boyuan and {Llanes-Estrada}, Felipe J. and {Loffredo}, Eleonora and {Long}, Oliver and {Lope-Oter}, Eva and {Lukes-Gerakopoulos}, Georgios and {Maggio}, Elisa and {Maggiore}, Michele and {Mancarella}, Michele and {Mapelli}, Michela and {Marchant}, Pablo and {Margiotta}, Annarita and {Mariotti}, Alberto and {Marriott-Best}, Alisha and {Marsat}, Sylvain and {Mart{\'\i}nez-Pinedo}, Gabriel and {Maselli}, Andrea and {Mastrogiovanni}, Simone and {Matos}, Isabela and {Melandri}, Andrea and {Mendes}, Raissa F.~P. and {Mendon{\c{c}}a Soares de Souza}, Josiel and {Mentasti}, Giorgio and {Mezcua}, Mar and {M{\"o}sta}, Philipp and {Mondal}, Chiranjib and {Moresco}, Michele and {Mukherjee}, Tista and {Muttoni}, Niccol{\`o} and {Nagar}, Alessandro and {Narola}, Harsh and {Nava}, Lara and {Navarro Moreno}, Pablo and {Nelemans}, Gijs},
        title = "{The Science of the Einstein Telescope}",
      journal = {arXiv e-prints},
         year = 2025,
        month = mar,
          eid = {arXiv:2503.12263},
        pages = {arXiv:2503.12263},
          doi = {10.48550/arXiv.2503.12263},
archivePrefix = {arXiv},
       eprint = {2503.12263},
 primaryClass = {gr-qc},
       adsurl = {https://ui.adsabs.harvard.edu/abs/2025arXiv250312263A}
}

@ARTICLE{Gongetal2021,
       author = {{Gong}, Yungui and {Luo}, Jun and {Wang}, Bin},
        title = "{Concepts and status of Chinese space gravitational wave detection projects}",
      journal = {Nature Astronomy},
         year = 2021,
        month = sep,
       volume = {5},
        pages = {881-889},
          doi = {10.1038/s41550-021-01480-3},
archivePrefix = {arXiv},
       eprint = {2109.07442},
 primaryClass = {astro-ph.IM},
       adsurl = {https://ui.adsabs.harvard.edu/abs/2021NatAs...5..881G}
}

@ARTICLE{Yao2025,
       author = {{Yao}, Yuhan and {Chornock}, Ryan and {Ward}, Charlotte and {Hammerstein}, Erica and {Sfaradi}, Itai and {Margutti}, Raffaella and {Kelley}, Luke Zoltan and {Lu}, Wenbin and {Liu}, Chang and {Wise}, Jacob and {Sollerman}, Jesper and {Alexander}, Kate D. and {Bellm}, Eric C. and {Drake}, Andrew J. and {Fremling}, Christoffer and {Gilfanov}, Marat and {Graham}, Matthew J. and {Groom}, Steven L. and {Hinds}, K.~R. and {Kulkarni}, S.~R. and {Miller}, Adam A. and {Miller-Jones}, James C.~A. and {Nicholl}, Matt and {Perley}, Daniel A. and {Purdum}, Josiah and {Ravi}, Vikram and {Rich}, R. Michael and {Rehemtulla}, Nabeel and {Riddle}, Reed and {Smith}, Roger and {Stein}, Robert and {Sunyaev}, Rashid and {van Velzen}, Sjoert and {Wold}, Avery},
        title = "{A Massive Black Hole 0.8 kpc from the Host Nucleus Revealed by the Offset Tidal Disruption Event AT2024tvd}",
      journal = {\apjl},
         year = 2025,
        month = jun,
       volume = {985},
       number = {2},
          eid = {L48},
        pages = {L48},
          doi = {10.3847/2041-8213/add7de},
archivePrefix = {arXiv},
       eprint = {2502.17661},
 primaryClass = {astro-ph.GA},
       adsurl = {https://ui.adsabs.harvard.edu/abs/2025ApJ...985L..48Y}
}

@ARTICLE{Khan2025LRD,
       author = {{Khan}, Fazeel Mahmood and {Davis}, Benjamin L. and {Macci{\`o}}, Andrea Valerio and {Holley-Bockelmann}, Kelly},
        title = "{Where Have All the Little Red Dots Gone? Supermassive Black Hole Binary Dynamics and Its Impact on Galaxy Properties}",
      journal = {\apjl},
         year = 2025,
        month = jun,
       volume = {986},
       number = {1},
          eid = {L1},
        pages = {L1},
          doi = {10.3847/2041-8213/adda4c},
archivePrefix = {arXiv},
       eprint = {2503.07711},
 primaryClass = {astro-ph.GA},
       adsurl = {https://ui.adsabs.harvard.edu/abs/2025ApJ...986L...1K}
}

@ARTICLE{duffell20,
       author = {{Duffell}, Paul C. and {D'Orazio}, Daniel and {Derdzinski}, Andrea and {Haiman}, Zoltan and {MacFadyen}, Andrew and {Rosen}, Anna L. and {Zrake}, Jonathan},
        title = "{Circumbinary Disks: Accretion and Torque as a Function of Mass Ratio and Disk Viscosity}",
      journal = {\apj},
         year = 2020,
        month = sep,
       volume = {901},
       number = {1},
          eid = {25},
        pages = {25},
          doi = {10.3847/1538-4357/abab95},
archivePrefix = {arXiv},
       eprint = {1911.05506},
 primaryClass = {astro-ph.SR},
       adsurl = {https://ui.adsabs.harvard.edu/abs/2020ApJ...901...25D}
}

@ARTICLE{franchini22,
       author = {{Franchini}, Alessia and {Lupi}, Alessandro and {Sesana}, Alberto},
        title = "{Resolving Massive Black Hole Binary Evolution via Adaptive Particle Splitting}",
      journal = {\apjl},
         year = 2022,
        month = apr,
       volume = {929},
       number = {1},
          eid = {L13},
        pages = {L13},
          doi = {10.3847/2041-8213/ac63a2},
archivePrefix = {arXiv},
       eprint = {2201.05619},
 primaryClass = {astro-ph.HE},
       adsurl = {https://ui.adsabs.harvard.edu/abs/2022ApJ...929L..13F}
}

@ARTICLE{siwek23,
       author = {{Siwek}, Magdalena and {Weinberger}, Rainer and {Hernquist}, Lars},
        title = "{Orbital evolution of binaries in circumbinary discs}",
      journal = {\mnras},
         year = 2023,
        month = jun,
       volume = {522},
       number = {2},
        pages = {2707-2717},
          doi = {10.1093/mnras/stad1131},
archivePrefix = {arXiv},
       eprint = {2302.01785},
 primaryClass = {astro-ph.HE},
       adsurl = {https://ui.adsabs.harvard.edu/abs/2023MNRAS.522.2707S}
}

@ARTICLE{bortolas21,
       author = {{Bortolas}, Elisa and {Franchini}, Alessia and {Bonetti}, Matteo and {Sesana}, Alberto},
        title = "{The Competing Effect of Gas and Stars in the Evolution of Massive Black Hole Binaries}",
      journal = {\apjl},
         year = 2021,
        month = sep,
       volume = {918},
       number = {1},
          eid = {L15},
        pages = {L15},
          doi = {10.3847/2041-8213/ac1c0c},
archivePrefix = {arXiv},
       eprint = {2108.13436},
 primaryClass = {astro-ph.HE},
       adsurl = {https://ui.adsabs.harvard.edu/abs/2021ApJ...918L..15B}
}

@ARTICLE{Partmann2024,
       author = {{Partmann}, Christian and {Naab}, Thorsten and {Rantala}, Antti and {Genina}, Anna and {Mannerkoski}, Matias and {Johansson}, Peter H.},
        title = "{The difficult path to coalescence: massive black hole dynamics in merging low-mass dark matter haloes and galaxies}",
      journal = {\mnras},
         year = 2024,
        month = aug,
       volume = {532},
       number = {4},
        pages = {4681-4702},
          doi = {10.1093/mnras/stae1712},
archivePrefix = {arXiv},
       eprint = {2310.08079},
 primaryClass = {astro-ph.GA},
       adsurl = {https://ui.adsabs.harvard.edu/abs/2024MNRAS.532.4681P}
}

@ARTICLE{hern1992,
       author = {{Hernquist}, Lars and {Ostriker}, Jeremiah P.},
        title = "{A Self-consistent Field Method for Galactic Dynamics}",
      journal = {\apj},
         year = 1992,
        month = feb,
       volume = {386},
        pages = {375},
          doi = {10.1086/171025},
       adsurl = {https://ui.adsabs.harvard.edu/abs/1992ApJ...386..375H}
}

@ARTICLE{KHB2025,
       author = {{Holley-Bockelmann}, Kelly and {Khan}, Fazeel and {Williams}, Isaiah and {Roth}, Jaelyn and {Rizzo Smith}, Michael and {Porter}, Kaitlin and {Bellovary}, Jillian and {Derdzinski}, Andrea and {Macci{\`o}}, Andrea},
        title = "{Handy Relation Between Binary Black Hole Merger Times and Host Galaxy Properties}",
      journal = {arXiv e-prints},
         year = 2025,
        month = aug,
          eid = {arXiv:2508.14253},
        pages = {arXiv:2508.14253},
          doi = {10.48550/arXiv.2508.14253},
archivePrefix = {arXiv},
       eprint = {2508.14253},
 primaryClass = {astro-ph.GA},
       adsurl = {https://ui.adsabs.harvard.edu/abs/2025arXiv250814253H}
}




\end{document}